\documentclass[nofootinbib,superscriptaddress,eqsecnum,tightenlines,11pt]{revtex4}

\usepackage{graphicx}
\usepackage{amsmath,amssymb,amsfonts,amsthm,stmaryrd,mathtools}
\usepackage{multirow,bigdelim}
\usepackage{dsfont}
\usepackage{color}
\usepackage{tikz}
\usepackage{subcaption}
\usetikzlibrary{shapes.geometric} 
\usepackage{hyperref}
 
\makeatletter
\newcounter{subsubsubsection}[subsubsection]
\renewcommand\thesubsubsubsection{\thesubsubsection .\@alph\c@subsubsubsection}
\newcommand\subsubsubsection{\@startsection{subsubsubsection}{4}{\z@}%
                                     {-3.25ex\@plus -1ex \@minus -.2ex}%
                                     {1.5ex \@plus .2ex}%
                                     {\centering\normalfont\small\textit}}
\newcommand*\l@subsubsubsection{\@dottedtocline{3}{10.0em}{4.1em}}
\newcommand*{\subsubsubsectionmark}[1]{}
\makeatother

\def\be{\begin{equation}}
\def\ee{\end{equation}}
\def\ba{\begin{eqnarray}}
\def\ea{\end{eqnarray}}
\def\bas{\begin{subequations}\begin{eqnarray}}
\def\eas{\end{eqnarray}\end{subequations}}

\def\SU{\text{SU}}
\def\SO{\text{SO}}

\def\nn{\nonumber}
\def\q{\  }

\def\dim{\text{dim}\,}
\def\im{\text{im}\,}
\def\ker{\text{ker}\,}

\def\c{\mathrm{c\,}\!}
\def\cs{\mathrm{c^2\,}\!}
\def\s{\mathrm{s\,}\!}

\begin{document}

\title{ The Degrees of Freedom of Area Regge Calculus: Dynamics, Non-metricity, and Broken Diffeomorphisms
}

\author{Seth K. Asante}
\email{sasante@perimeterinstitute.ca}
\affiliation{Perimeter Institute for Theoretical Physics, 31 Caroline Street North, Waterloo, ON, N2L 2Y5, Canada}
\affiliation{Department of Physics and Astronomy, University of Waterloo, Waterloo, ON, N2L 3G1, Canada}

\author{Bianca Dittrich}
\email{bdittrich@perimeterinstitute.ca}
\affiliation{Perimeter Institute for Theoretical Physics, 31 Caroline Street North, Waterloo, ON, N2L 2Y5, Canada}

\author{Hal M. Haggard}
\email{haggard@bard.edu}
\affiliation{Physics Program, Bard College, 30 Campus Road, Annondale-On-Hudson, NY 12504, USA}
\affiliation{Perimeter Institute for Theoretical Physics, 31 Caroline Street North, Waterloo, ON, N2L 2Y5, Canada}

\begin{abstract}
Discretization of general relativity is a promising route towards quantum gravity. Discrete geometries have a finite number of  degrees of freedom and can mimic aspects of quantum geometry. However, selection of the correct discrete freedoms and  description of their dynamics has remained a challenging problem.  We explore classical area Regge calculus, an alternative to standard Regge calculus where instead of lengths, the areas of a simplicial discretization are fundamental. There are a number of surprises: though the equations of motion impose flatness we show that diffeomorphism symmetry is broken for a large class of area Regge geometries. This is due to degrees of freedom not available in the length calculus. In particular, an area discretization only imposes that the areas of glued simplicial faces agrees; their shapes need not be the same. We enumerate and characterize these non-metric, or `twisted', degrees of freedom and provide tools for understanding their dynamics. The non-metric degrees of freedom also lead to fewer invariances of the area Regge action---in comparison to the length action---under local changes of the triangulation (Pachner moves). This means that invariance properties can be used to classify the dynamics of spin foam models. Our results lay a promising foundation for understanding the dynamics of the non-metric degrees of freedom in loop quantum gravity and spin foams. 

\end{abstract}

\maketitle

\section{Introduction}\label{introduction}

In  general relativity the properties of gravity are encoded into the geometry of spacetime. One therefore expects a theory of quantum gravity to be based on quantum geometries.  By now we have many different notions of quantum geometry and which is best suited for quantum gravity is unclear.

A critical question, especially in discretizing general relativity, is which geometric variables provide the fundamental degrees of freedom.  Regge calculus \cite{Regge} is a beautiful geometric framework that provides a discretization of general relativity based on flat simplices. The Regge action agrees with the exact value of the continuum action (including boundary terms)  for these flat building blocks.  One can also work with homogeneously curved constituents \cite{Improved,NewRegge}, which on the quantum level often involves a quantum deformation of the underlying structure group 
\cite{TuraevViro:1992,Crane:94,Major:96,Barrett:03,Dupuis:14,Bonzom:14,Haggard:15,Haggard:16,Haggard:162,Haggard:18,Dittrich:17,Dittrich:172}.
Even in Regge calculus the choice of fundamental variables is not unique. Regge introduced the discretization using the simplex lengths as variables, but comparison between three- and four-dimensional quantum gravity suggests that areas may provide better discretization variables in 4D \cite{Rovelli:93}. 
Thus other versions of Regge calculus have been introduced, in particular area Regge calculus \cite{AreaRegge1}  and area--angle Regge calculus \cite{AreaAngle}. These are  equivalent to length Regge calculus if one  implements constraints that ensure that the configurations considered arise from a consistent length assignment \cite{AreaReggeConstraints, AreaAngle}. Weakening these constraints or not implementing them  at all leads, however, to a different dynamics based on a configuration space of generalized discrete geometries. 

Area variables appear to be more fundamental when one favors a gauge formulation of gravity (e.g. with gauge group $\SU(2)$ or $\SO(3,1)$, \cite{AshtekarVariables}). To characterize the curvature of a spacetime one uses (Ashtekar) connection variables, which are integrated along paths and exponentiated to holonomies. The natural conjugated variables are electric (or triad) fields, which in the $(3+1)$-dimensional case are integrated over two-dimensional surfaces---from these variables one constructs the areas of these surfaces
\cite{DiscreteGeom1,DiscreteGeom2}.
The same structure of canonically conjugated variables appears in lattice gauge field theories. 

In  constructions of a phase space for (3+1)--dimensional simplicial geometries the simplest choice of conjugated variables is also given by the areas of the triangles and the dihedral angles hinging on them \cite{DittDiffReview,DittRyan1}. The dihedral angles encode the extrinsic curvature and the areas the intrinsic curvature.  In contrast, if one chooses lengths as configuration variables, the conjugated observables are more complicated combinations of the dihedral angles and area-length derivatives \cite{DittrichHoehn1}.

In spite of these motivating arguments, questions about area and area-angle Regge calculus abound. Even for a small simplicial complex, the areas greatly outnumber the edge lengths. What is the nature of these extra degrees of freedom? In particular, there are multidimensional families of area configurations that have no corresponding description in the length Regge calculus. What are these `non-metric' area Regge configurations? Do they have a correspondent in continuum general relativity? In addition, studies of length Regge calculus have shown that length geometries carrying curvature break the diffeomorphism symmetry of the continuum theory \cite{BahrDittrich1}. Thus one is led to wonder: Does area Regge calculus break diffeomorphism symmetry? Under what conditions? Most importantly, can we understand the dynamics of these degrees of freedom? These are the questions that we take up in this paper. In particular, we are interested in the light that answering them will shed on related questions in quantum gravity, and in particular loop quantum gravity.

Loop quantum gravity provides a highly developed theory of quantum geometries. On the classical level  various phase space descriptions of simplicial geometries \cite{DittRyan1, SimplPhaseSpace1,SimplPhaseSpace2,SimplPhaseSpace3}
have been constructed.  In the quantum theory rigorous representations of geometric observables as operators on continuum Hilbert spaces  are available 
\cite{ObsRep1,ObsRep2,ObsRep3}.
And in the covariant theory spin foam amplitudes 
\cite{BarrettCrane, NewSFM1,NewSFM2,NewSFM3,NewSFM4,NewSFM5}
describe a dynamics for these quantum geometries.  The quantum geometries can be analyzed in a semiclassical limit 
\cite{SemCl0,SemCl1,SemCl1.5,SemCl2,SemCl2.5,SemCl3,SemCl4,SemCl5,vectorgeom3},
which here means that the discrete areas become large in comparison to the Planck area.  These works show that  the Regge action \cite{Regge}  emerges in this semiclassical limit. 

Indeed, area Regge calculus was first suggested as a description of the classical limit of the first four-dimensional spin foam model for gravity, the Barrett--Crane model \cite{Rovelli:93}.  It was quickly noted \cite{AreaRegge1}, however, that area Regge gravity does {\it not} describe general relativity. One reason is that, as we will see, the equations of motion impose flatness. The second reason is that  in three- and four-dimensional triangulations there are typically far more triangles than edges and thus far more area variables than length variables. This means that using (unconstrained) area variables, one describes a much bigger configuration space than that provided by (piecewise) simplicial geometries.  


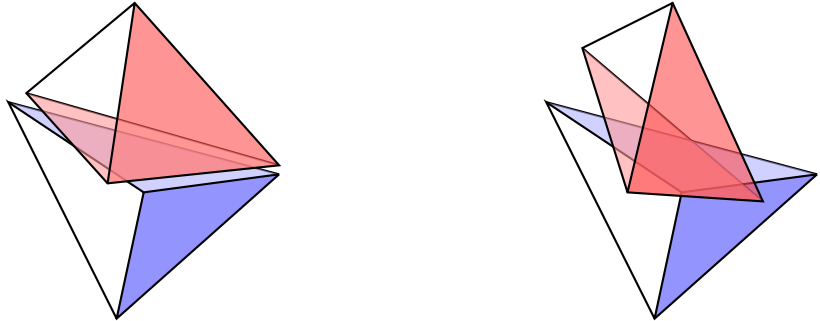
\begin{figure}[t]
\begin{tikzpicture}[scale = 1.2]
   \draw[thick] (-1.2,0)--(1.6,-0.8) ;
   \draw[thick]  (1.6,-0.9)--(-1.4,.-0.1) ;
   \path[fill= blue!70,opacity= 0.6] (-0.2,-2.5)--(0.1,-1.1)--(1.6,-0.9);
    \path[fill= blue!30,opacity= 0.6] (-1.4,-0.1)--(0.1,-1.1)--(1.6,-0.9);
   \draw[thick] (0.1,-1.1)--(-0.2,-2.5) (1.6,-0.9)--(0.1,-1.1)--(-1.4,.-0.1)--(-0.2,-2.5)--(1.6,-0.9);
    \path[fill= red!70,opacity= 0.6] (0,1)--(-0.3,-1)--(1.6,-0.8);
   \path[fill= red!40,opacity= 0.6] (-1.2,0)--(-0.3,-1)--(1.6,-0.8);    
   \draw[ thick] (0,1)--(-0.3,-1) (0,1)--(-1.2,0)--(-0.3,-1)--(1.6,-0.8)--(0,1)--cycle;
\end{tikzpicture}
\hspace{.2\textwidth} 
\begin{tikzpicture}[scale = 1.2]
   \draw[thick] (-1,.5)--(1,-1.2) ;
   \draw[thick]  (1.6,-0.9)--(-1.4,.-0.1) ;
   \path[fill= blue!70,opacity= 0.6] (-0.2,-2.5)--(0.1,-1.1)--(1.6,-0.9);
    \path[fill= blue!30,opacity= 0.6] (-1.4,-0.1)--(0.1,-1.1)--(1.6,-0.9);
   \draw[thick] (0.1,-1.1)--(-0.2,-2.5) (1.6,-0.9)--(0.1,-1.1)--(-1.4,.-0.1)--(-0.2,-2.5)--(1.6,-0.9);
   \path[fill= red!40,opacity= 0.6] (-1,.5)--(-0.5,-1.1)--(1,-1.2);    
       \path[fill= red!70,opacity= .6] (0,1)--(-0.5,-1.1)--(1,-1.2);
   \draw[ thick] (0,1)--(-0.5,-1.1) (0,1)--(-1,.5)--(-0.5,-1.1)--(1,-1.2)--(0,1)--cycle;
\end{tikzpicture}
\caption{ Two examples of the generalized simplicial geometries considered in quantum gravity. In both cases the pairs of tetrahedra are glued along their pale shaded faces and the areas of these triangles agree. On the left the shape mismatch is mild, while on the right it is more extreme. }\label{ShapeMismatch}
\end{figure}

More recently arguments have been put forward that the Barrett--Crane model cannot lead to the dynamics of general relativity \cite{Alesci}, see also the discussion \cite{OritiAboutBC}. This motivated the introduction of `new' models \cite{NewSFM1,NewSFM2,NewSFM3,NewSFM4,NewSFM5}, whose boundary Hilbert spaces match those of canonical loop quantum gravity. However, these new models, and more generally (canonical) loop quantum gravity, describe a class of generalized simplicial geometries. These generalized geometries identify a well-defined length geometry for each tetrahedron of the canonical formulation (or for each 4-simplex of the covariant formulation). But, in gluing the tetrahedra or simplices together, it turns out that although the areas of pairs of identified triangles match, this need not be the case for the shape of each glued triangle. Figure \ref{ShapeMismatch} illustrates these generalized geometries for two pairs of tetrahedra. Constraints that impose the matching of these shapes have been worked out in \cite{AreaAngle} and are known as gluing conditions or shape matching constraints.
 
   Intriguingly, shape matching constraints are also {\it not} automatically implemented in a classical version of loop quantum gravity, that is,  in the description of the phase space of $(3+1)$-dimensional simplicial geometries \cite{DittRyan1}. This has been linked, in \cite{DittRyan2}, to the much discussed question of whether one should implement, in addition to the primary simplicity constraints,  secondary simplicity constraints in spin foam models or not
   \cite{ CanonQuant1,CanonQuant2,CanonQuant3}.
 References \cite{DittRyan3} and \cite{GeillerNouiBarb}  show that the enlarged space of simplicial geometries can also  explain the appearance of the Barbero-Immirzi parameter in loop quantum gravity.\footnote{That is, the fact that the Barbero-Immirzi parameter appears in the spectrum of geometric observables, despite the fact that classically it only parametrizes a canonical transformation of the theory's variables.}

The work \cite{SpezFreid} described these same generalized simplicial geometries in a  phase space parametrization that included a `twisting angle' and provided a more direct derivation from canonical loop quantum gravity, using the phase space associated to the cotangent bundle $T^*\SU(2)$ of each triangle. (See \cite{BonzomAreaAngle} for a derivation of the twisting angles from spin foams.) This led to the term `twisted' geometries, which is now the most common name for this class of generalized simplicial geometries.\footnote{ See also \cite{FreidZiprick} for an alternative interpretation of twisted geometries in terms of `spinning' geometries.} Indeed a link to twistors has been proposed in \cite{SpezFreid2,LivineSpezTamb}, and extended further in \cite{WielandSpez}. A proposal for a 4-simplex action for twisted geometries beyond the
shape-matched sector appears in \cite{BanburskiEtAlPachner}.  Secondary simplicity
constraints do arise in this context \cite{SpezialeFabio} and, intriguingly, only admit solutions in the sector of shape matched configurations. On the other hand, it was shown in \cite{HalWolfgang} that there exist alternative definitions of the torsionless condition that can be nicely solved for shape mismatched configurations. This highlights
the question of what kind of dynamics one can attribute to these generalized geometries and is one of our central themes. 

While \cite{HalWolfgang} provides a construction for a Levi--Civita connection for the twisted geometries,  in a sense, they identify an exponentiated version of the symmetric part of a more general connection. There is a left-over part of the connection's holonomy that acts within the plane of the triangle along which two tetrahedra are glued. This can be understood as the non-symmetric part and therefore as describing torsion degrees of freedom. This provides yet another alternative interpretation for the additional degrees of freedom that appear for the generalized or twisted geometries. 

 In this paper we will refer to these degrees of freedom as non-metric, in the sense that they extend the space of simplicial, piecewise linear and piecewise flat (or homogeneously curved) geometries beyond a length description. 

One has thus reached quite a detailed understanding of the generalized space of geometries that underlies loop quantum gravity. So far this understanding is strictly on the kinematical level and there is not a clear understanding of what kind of dynamics the spin foam models prescribe for the additional degrees of freedom present in these non-metric geometries.  This understanding is necessary to clarify whether the current spin foam models can describe the dynamics of general relativity, or whether additional constraints need to be added to suppress the non-metric degrees of freedom. Again, see \cite{SpezialeFabio} for a first analysis of this question.

Even on the classical level, the dynamics (and kinematics) of area Regge calculus remains poorly understood \cite{BarrettRefWaves,Wainwright,Neiman}. But, area Regge calculus does provide a dynamics for the non-metric degrees of freedom, which also appear in spin foams. The action of area Regge calculus numerically coincides with that of length Regge calculus (on configurations which can be matched to each other), and as we have noted above the Regge action (in variables that include areas) appears as the classical limit of Barrett-Crane-type spin foam models. In fact, area Regge calculus is still the best candidate for the classical limit of the Barrett-Crane model.  The semiclassical analysis of the newer spin foam models feature  the appearance of dominating saddle points that describe so-called vector geometries \cite{vectorgeom, vectorgeom2,vectorgeom3}. Additional shape mismatched configurations  and generalized Regge actions appear also at the saddle point of non-simplicial spin foams, as shown in \cite{HyperCubeCoarse} for regular hypercuboids and in \cite{vectorgeom3} more generally.
While area Regge calculus will shed light on the dynamics of all these spin foams, we expect that a better candidate for their classical limit is area-angle Regge calculus \cite{vectorgeom3}, which includes the three-dimensional dihedral angles as independent variables \cite{AreaAngle}.

Here we aim at a broader understanding of the possible dynamics of such generalized geometries arising from loop quantum gravity.  This can provide effective descriptions for quantum gravity models, which will help to understand their dynamics and to improve the models. We also aim to provide a foundation  and tools for future studies of the dynamics of area-angle Regge calculus.

Theories in which areas are the fundamental variables are also interesting in a wider context. The investigations of area metrics in the continuum by Schuller et al \cite{Schuller:2005yt} are motivated  by string theory. Recently  it has been suggested that quantum gravity and a notion of quantum geometry can be rebuilt from the entanglement structure of (possibly matter) quantum fields \cite{Raamsdonk,Raamsdonk2}. In particular  Ryu and Takayanagi  \cite{RyuTak} propose that in a holographic setup the entanglement of the dual boundary field can be used to measure the areas (in (3+1) dimensions) of surfaces extending into the bulk.  Thus, also here, areas appear to be more fundamental. In some sense areas are more natural than length in $(3+1)$ dimensions  as the flow of a vector field through a surface can be used to measure the area of this surface.

In this work we will therefore revisit area Regge calculus. In Section \ref{Sec:areaRegge}, we  address an ambiguity problem that arises in the original formulation of the theory, which makes the action ill-defined for configurations with right angles. We circumvent this problem by constructing a first order formulation. In Section \ref{LinTheory} we analyze certain aspects of the dynamics of linearized area Regge calculus. Here we consider, in particular, setups that are helpful in distinguishing between length-Regge-type dynamics and area-type dynamics in spin foams. To this end we consider configurations that describe Pachner moves, that is, local changes in the triangulation in Section \ref{Pachnermoves}. We will indeed see that length and area Regge calculus behave differently under these Pachner moves.   We provide a canonical analysis of the dynamics using tent moves in Section \ref{Tentmoves}. Tent moves also allow a comparison of the counting of (propagating or physical) degrees of freedom between length and area Regge calculus. We will see that area Regge calculus has generically more propagating degrees of freedom than length Regge calculus, and that the additional degrees of freedom can be matched to specific variables describing the non-matching of the shapes of (glued) triangles. 

As part of this analysis of the dynamics we will also see that linearized area Regge calculus---at least on so-called metric backgrounds---features gauge symmetries that will lead to (first class) constraints for the linearized theory.  These gauge symmetries describe the displacement of vertices in the triangulation and can be understood as discrete remnants of diffeomorphism symmetry \cite{DittDiffReview}. That is, solutions remain invariant under certain changes of the areas of triangles adjacent to a vertex. The same kind of gauge symmetries appear for linearized length Regge calculus \cite{WilliamsR}---but only for flat backgrounds. Indeed it has been shown in \cite{BahrDittrich1} that changing to a (background) solution with curvature will break these symmetries.  Diffeomorphism symmetry is a fundamental symmetry of general relativity that one would like to preserve during quantization.\footnote{The breaking of diffeomorphism symmetry has strong implications for the canonical formulation, which are linked to the appearance of anomalies for the constraint algebra, see the discussions in \cite{GambiniPullin, BahrDittrich1, DittrichHoehn1}.} These observations about diffeomorphism breaking in discrete geometric theories have motivated a coarse graining and renormalization program \cite{Ditt12Review,CoarseRenorm1,CoarseRenorm2,CoarseRenorm3,CoarseRenorm4,CoarseRenorm5,CoarseRenorm6,CoarseRenorm7,HyperCubeCoarse} 
aimed at finding improved discrete actions and spin foam models that feature a more complete version of diffeomorphism symmetry. 

In Section \ref{diffbreaking} we show that non-metricity (torsion) also breaks diffeomorphism symmetry.  In particular, we show that certain three-dimensional dihedral angles can be used to capture the peculiar non-metricity of area Regge calculus and to parametrize the extent of the diffeomorphism breaking. We explore the implications of all of these findings in the discussion, Section \ref{discussion}.

\section{Area Regge calculus}\label{Sec:areaRegge}

In  area Regge calculus, \cite{AreaRegge1}, one assumes a four--dimensional triangulation $\Delta$  and associates an area variable $A_t$ to each triangle $t$. The action is then  
\ba\label{action1}
S=\sum_{t\in \Delta} A_t \epsilon_t(A_{t'})  \q ,
\ea
where 
\ba\label{deficit}
\epsilon_t = 2\pi - \sum_{\sigma \supset t} \theta^\sigma_t( \{A_{t'}\}_{t'\subset \sigma})
\ea
is the deficit angle at this triangle.  The deficit angle is computed from the 4D  (internal) dihedral angles $\theta_t^\sigma$, which are the angles between the two sub-tetrahedra of the four-simplex $\sigma$ that share the triangle $t$.   When the complex has a boundary, we adopt as boundary conditions fixed area variables. The boundary term is analogous to the Gibbons-Hawking-Hartle-Sorkin boundary term \cite{HartleSorkin} of length Regge calculus
\ba
S_{\rm bdry}=\sum_{t\in \partial \Delta} A_t (\pi - \sum_{\sigma\subset t} \theta^\sigma_t( \{A_t'\}_{t'\subset \sigma})) \q .
\ea

The $\theta_t^\sigma$ are uniquely determined from the (flat) geometry of the simplex $\sigma$, which is defined by its 10 edge lengths. Eq. (\ref{action1}), however, requires the dihedral angles as functions of the 10 triangle areas of the simplex.  Unfortunately, the 10 areas of a four-simplex do not uniquely determine the 10 length variables. This is due to the fact that the areas are quadratic functions of the edge lengths. A particular example of this ambiguity is the ``Tuckey--configuration," which has all edge lengths equal to 1 except for one edge with length $\sqrt{b}$, \cite{AreaRegge1}.  For both the values $b$ and $4-b$, triangles sharing the latter edge have equal areas. Nonetheless, locally in configuration space and away from right angle configurations, one can invert the areas for the lengths. There are in general multiple roots for each length and these roots coalesce at the right angle configurations. At these configurations  the Jacobian $\partial A_t/\partial l_e$ is not invertible.

Let us therefore assume that the action (\ref{action1}) can be well defined (by e.g. selecting roots) in a certain region of configuration space. The equations of motion impose
\ba
\epsilon_t=0 \q ,
\ea
that is, flatness of the simplicial complex. This is due to the Schl\"afli identity \cite{Schlafli:1858} (for a modern symplectic proof see \cite{HaggardSch:2015})
\ba\label{schlafli}
\sum_{t\subset \sigma} 
A_t  \, \delta \theta_t^\sigma \,=\,0 \q ,
\ea
which holds for arbitrary variations  $\delta \theta_t^\sigma$ of the dihedral angles in a simplex $\sigma$ and, in effect, leads to a vanishing of the variations of the deficit angles  $\sum_t A_t \,\delta \epsilon_t=0$.

The equations of motion  impose flatness. To cure the problem with the ambiguities of the action we will consider first order Regge calculus. An alternative, presented in section \ref{curvedRegge},  is to replace flat simplices by simplices with homogeneous curvature. This framework is useful for modeling spacetimes with a cosmological constant. We will  work here mostly with the flat simplex version.

\subsection{First order Area--Regge calculus}

To circumvent the problem of finding the dihedral angles as functions of the areas we will treat these dihedral angles $\theta_t^\sigma$ as independent variables. This amounts to a first order area Regge calculus. First order frameworks for the standard length Regge calculus were defined by Barrett \cite{BarrettFO}  for the flat case and in \cite{NewRegge} for the case of homogeneously curved simplices.

The mechanism behind our first order formulation is the same as in \cite{BarrettFO}, with the important difference that here the equations of motion impose flatness. On a given simplex the full set of dihedral angles and areas provides more data than necessary to determine the geometry of the simplex.  Hence these variables cannot be specified completely independently, or the dihedral angles might not be compatible with the areas. Consistency of these data will be imposed by an equation of motion that follows from the variation of the dihedral angles.

The dihedral angles of a (flat) simplex are also not independent. Define the angle Gram matrix 
\ba
G^\sigma_{ij} \,:=\, -\cos(\theta_{ij}^\sigma) \quad \text{for} \quad i\neq j, \quad \text{and} \quad \q\q G^\sigma_{ii}=1 \q,
\ea
where $i,j \in \{1,\ldots,5\}$ label the five vertices of the simplex $\sigma$ and $\theta^\sigma_{ij}$ is the dihedral angle opposite the edge connecting vertices $i$ and $j$. Then the dihedral angles must satisfy the constraint that the determinant of the angle Gram matrix $G^\sigma$ vanishes.   In Appendix \ref{SimplexK} we prove this claim and give a general structural characterization of the Gram matrix. 

As in \cite{BarrettFO}, we impose this constraint on each simplex using a Lagrange multiplyer $\Lambda_\sigma$. The action is
\ba\label{action2}
S=\sum_{t\in \Delta} A_t (2\pi - \sum_{\sigma \supset t} \theta^\sigma_t)\,+\, \, \sum_\sigma \Lambda_\sigma  \det G^\sigma  \q .
\ea 
This action leads to the equations of motion
\ba\label{EOM2}
&&\delta A_t:  \quad \quad \epsilon_t= 2\pi - \sum_{\sigma \supset t} \theta^\sigma_t\,=\,0  \q,\nn\\
&&\delta \Lambda_\sigma:\quad \quad  \det G^\sigma=0 \q, \nn\\
&& \delta \theta_t^\sigma:\quad \quad  A_t= \Lambda_\sigma \,  \frac{\partial \det G^\sigma}{\partial \theta^\sigma_t}  \q .
\ea
In Appendix \ref{SimplexK} we find the derivative of the determinant of the angle Gram matrix 
\ba\label{detG}
\frac{\partial \det G^\sigma}{\partial \theta^\sigma_{ij}}  \,=\,   \left[c V_iV_j\right]\!\!(\theta_{kl}^\sigma)\, \, \sin \theta^\sigma_{ij} \,=\, \left[c'  V_{ij}\right]\!\!(\theta_{kl}^\sigma) \q,
\ea
where $V_i$ is the volume of the tetrahedron obtained by removing from $\sigma$ the vertex $i$ and $V_{ij}$ is the area of the triangle obtained by removing vertices $i$ and $j$. The coefficients $c$ and $c'$ are  defined in Appendix \ref{SimplexK} and only depend on the full simplex. They are dimensionful, as we have a dimensionless quantity on the left hand side of (\ref{detG}). On the right hand side we have made the dependencies on the angles $\theta^{\sigma}$ explicit, but, as the dihedral angles cannot determine the scale of the simplex, it is only the combined quantities in square brackets that are well defined functions of the dihedral angles.

Thus, the last equation of motion in (\ref{EOM2}) does, in fact, impose that the dihedral angles are compatible with the areas. This will be more manifest in the homogeneously curved case considered below. The second equation of motion imposes that the dihedral angles come from a simplex.  In general (away from right angles) the last two equations of  (\ref{EOM2})  can be solved for the dihedral angles and the Lagrange multiplier in terms of the areas. This can be achieved locally on each simplex.

\subsection{ Area Regge calculus for homogeneously curved simplices}
\label{curvedRegge}

A second approach is to use homogeneously curved simplices. For a curved four-simplex, the curvature scale breaks the overall scaling symmetry of the flat case and both the edge lengths and the areas can be expressed as functions of the dihedral angles. Actions that impose the dynamics for length Regge calculus with homogeneously curved simplices have been investigated in \cite{NewRegge}, see also \cite{TuraevViro:1992,Crane:94,Major:96,Barrett:03,Dupuis:14,Bonzom:14,Haggard:15,Haggard:16,Haggard:162,Haggard:18,Dittrich:17,Dittrich:172} for spin foam models and loop  gravity techniques based on homogeneously curved simplices. 

  Homogeneously curved simplices allow for a simple solution to the dynamics of length Regge calculus with a cosmological constant. 
  While in the flat case it is intricate to arrive at expressions for the Hessian of the action, they are immediate in the constant curvature case. 
  This will allow us to easily find a second order formulation as well. In the first order action it will not be necessary to introduce a Lagrange multiplier as the cosmological constant and curved Schl\"afli identity address the associated issues completely. 

The first order action with cosmological constant $\Lambda = 3 \kappa$ is 
\be\label{actioncurved}
S_c = \sum_{t \in \Delta} A_t(2\pi - \sum_{\sigma \supset t} \theta_{t}^{\sigma})+ 3 \kappa \sum_{\sigma} V_{\sigma}(\theta^{\sigma})\q . 
\ee
Here $ V_{\sigma}(\theta^{\sigma})$ is the volume of the simplex $\sigma$, viewed as a function of all of its dihedral angles. The curved Schl\"afli identity gives the variation of this volume under an arbitrary variation of the dihedral angles
\be
3 \kappa \delta V_{\sigma} = \sum_{t \subset \sigma} A_{t} \delta \theta_{t}^{\sigma} \q. 
\ee 
Note that this properly reduces to the flat Schl\"afli identity, Eq. \eqref{schlafli}, in the $\kappa \rightarrow 0$ limit. Using the Schl\"afli identity, we have 
\be
\label{curvedSchlafli}
\frac{\partial V_{\sigma}}{\partial \theta_{t}^{\sigma}}   = \frac{1}{3 \kappa} A_{t}(\theta^{\sigma})
\ee
and the equations of motion follow immediately
\ba
\delta A_t:& \qquad \epsilon_{t} = 0 \nn \\
\delta \theta_{t}:& \qquad -A_t +A_t(\theta^{\sigma}) = 0.
\ea
The second of these equations imposes the area agreement, namely, that the independent area variables $A_{t}$ agree with the areas determined by the dihedral angles $A_{t}(\theta^{\sigma})$.

In anticipation of the more complicated linearization of the next section, we conclude this section by briefly considering linearization of the constant curvature area Regge calculus. 
First, split the variables into background values and perturbations
\ba
A_t &=& A_t^0+a_t \q, \nn\\
\theta^\sigma_t &=& (\theta^\sigma_t)^0 + \vartheta^\sigma_t \q .
\ea

The quadratic part of the action can be written as a contribution for each simplex 
\be
S^{(2)}_c = \sum_{\sigma} S_{c \sigma}^{(2)}
\ee
with the quadratic simplex action
\be
S_{c\sigma}^{(2)} \,=\,  - \sum_{t\subset \sigma}  a_t \vartheta^\sigma_t  \,+\,  \frac{3 \kappa}{2} \sum_{t,t' \subset \sigma} \vartheta^\sigma_t  \frac{\partial^2 S_c}{\partial \theta^\sigma_t \partial \theta^\sigma_{t'}}  \vartheta^\sigma_{t'}  \,=\,  - \sum_{t\subset \sigma}  a_t \vartheta^\sigma_t  \,+\,  \frac{3 \kappa}{2} \sum_{t,t' \subset \sigma} \vartheta^\sigma_t  \frac{\partial A_{t'}(\theta^{\sigma})}{\partial \theta^\sigma_t }  \vartheta^\sigma_{t'} \q ,
\ee
where in the second equality we have used the curved Schl\"afli identity \eqref{curvedSchlafli} again. The inverse function theorem guarentees that, at least locally, the Hessian \be
H_{t t'}^{\sigma} = \frac{\partial^2 S_c}{\partial \theta^\sigma_t \partial \theta^\sigma_{t'}} = \frac{\partial A_{t'}(\theta^{\sigma})}{\partial \theta^\sigma_t }
\ee
is invertible and has inverse
\be
(H^{\sigma})^{-1}_{tt'} = \frac{\partial \theta^\sigma_t}{\partial A_{t'}} \q. 
\ee
These explicit forms allow us to integrate out the $\vartheta^\sigma_{t}$ to obtain the second order  linearized action
\be
S^{(2)}_c = -\frac{1}{3 \kappa} \sum_{t,t' \subset \sigma} a_t (H^{\sigma})^{-1}_{t t'} a_{t'} \q.
\ee

While we have found $A_{t}(\theta^{\sigma})_{t'}$ analytically, see Appendix \ref{curvedanalytics}, inverting these functions to $\theta^{\sigma}_{t'}(A_t)$ has so far remained impractical.  Nonetheless the Hessians are straightforward to work with numerically and our results on the homogeneously curved case are found this way.

\section{Linearized theory}\label{LinTheory}

\subsection{The linearized  simplex action}\label{LinSim}

To analyze the symmetries of the theory and distinguish between physical and gauge degrees of freedom we will consider the linearized theory for flat simplices. (We have also tested some features for the curved simplex case, which we will comment on throughout.) 
To this end  we will assume a background given by a flat and metric configuration. Flatness is imposed by the equations of motion. Metric means that all the areas are determined from a consistent set of length variables. Section \ref{diffbreaking} considers more general backgrounds, and we will see that this is a very special choice of background with an enhanced symmetry content. This is closely analogous to the special character of  flat backgrounds in length Regge calculus \cite{WilliamsR,BahrDittrich1}.

Let us also consider the possibility of a boundary $\partial \Delta$ of the triangulation $\Delta$. As mentioned in Section \ref{Sec:areaRegge} we will adopt the Gibbons--Hawking--Hartle--Sorkin boundary term of standard Regge calculus and keep the areas on the boundary fixed:
\ba\label{action3}
S=\sum_{t\in \partial\Delta} A_t (\pi - \sum_{\sigma \supset t} \theta^\sigma_t)\,+\,\sum_{t\in \overset{\circ}\Delta} A_t (2\pi - \sum_{\sigma \supset t} \theta^\sigma_t)\,+\, \, \sum_\sigma \Lambda_\sigma  \det G^\sigma \q, 
\ea
here $\overset{\circ}\Delta$ denotes the bulk triangulation. Again we split the variables into background plus perturbations
\ba
A_t &=& A_t^0+a_t \q ,\nn\\
\Lambda^\sigma &=& \Lambda_\sigma^0  + \lambda_\sigma \q, \nn\\
\theta^\sigma_t &=& (\theta^\sigma_t)^0 + \vartheta^\sigma_t \q .
\ea

The quadratic part of the action can be written as a contribution for each simplex (with a boundary there might also be a linear boundary term for the expanded action)
\ba
S^{(2)}\,=\, \sum_\sigma S_\sigma
\ea
with
\ba
S_\sigma\,=\,  - \sum_{t\subset \sigma}  a_t \vartheta^\sigma_t  \,+\, \lambda_\sigma  \sum_{t\subset \sigma} \frac{\partial \det G^\sigma}{\partial \theta^\sigma_t}\vartheta^\sigma_t \,+\, \frac{\Lambda^0_\sigma}{2} \sum_{t,t' \subset \sigma} \vartheta^\sigma_t  \frac{\partial^2 \det G^\sigma}{\partial \theta^\sigma_t \partial \theta^\sigma_{t'}}  \vartheta^\sigma_{t'}  \q .
\ea
This can be written as
\ba
S_\sigma\,=\,  - \sum_{\tilde{t}\supset \sigma}  a_{\tilde{t}} \vartheta^\sigma_{\tilde{t}}  \,+\,  \frac{1}{2} \sum_{\tilde t, \tilde t' }  \vartheta^\sigma_{\tilde t} H^\sigma_{\tilde t \tilde t'} \vartheta^\sigma_{\tilde t}  \q, 
\ea
where we have introduced the extended index $\tilde t =(0,t)$ with $\vartheta^\sigma_{\tilde t=0}=\lambda_\sigma$ and $a_{\tilde t=0}=0$. The matrix $(H^\sigma)_{\tilde t \tilde t'}$ is given by 
\ba
 H^\sigma_{00}=0 \q , \q \q H^\sigma_{0t}&=& H_{t0} \,=\, \frac{\partial \det G^\sigma}{\partial \theta^\sigma_t} \, ,\nn\\
\text{and} \qquad H^\sigma_{tt'}&=&\Lambda^0_\sigma \frac{\partial^2 \det G^\sigma}{\partial \theta^\sigma_t \partial \theta^\sigma_{t'}}  \q .
\ea
Generically (away from right angles) this matrix can be inverted, that is, we can integrate out the dihedral angles and the $\lambda$ variables. This gives an effective simplex action expressed in terms of the area perturbations alone
\ba
S^\sigma_a \,=\,    -\frac{1}{2} \sum_{\tilde{t},\tilde{t}' \subset \sigma}  a_{\tilde{t}} \, ((H^\sigma)^{-1})_{\tilde{t} \tilde{t}'} \,a_{\tilde{t}'} = -\frac{1}{2} \sum_{t,t' \subset \sigma}  a_t \, ((H^\sigma)^{-1})_{tt'} \,a_{t'} \q,
\ea
where $((H^\sigma)^{-1})_{tt'} $ are the $tt'$-components of the inverse of the full matrix $H_{\tilde t \tilde t'}^\sigma$.  

Leveraging the bordered structure of $H_{\tilde t \tilde t'}^\sigma$, the matrix $((H^\sigma)^{-1})_{tt'} $ satisfies  the condition
 \ba
0&=& \sum_{t\in \sigma} \frac{\partial \det G^\sigma}{\partial \theta^\sigma_t} ((H^\sigma)^{-1})_{tt'} \nn\\
 &=& \sum_{t \in \sigma}  \, c'A_t(\theta^\sigma) \,\,((H^\sigma)^{-1})_{tt'}     ,
 \ea
 where in the second line we used  \eqref{detG}.  We are hence looking for a set of vectors $\{v^t\}_{t'}$ which are orthogonal to the vector with components the areas, $w^t=A_t$. The Schl\"afli identity does indeed provide such vectors: $v^t = \delta \theta_t^\sigma$ for any variation $\delta$. Additionally $ ((H^\sigma)^{-1})_{tt'}$ is a symmetric matrix, which suggests 
\ba\label{HessFlat}
((H^\sigma)^{-1})_{tt'}  = \frac{\partial \theta^\sigma_t}{\partial A_{t'}} \q .
\ea
This is confirmed by the fact that the same Hessian  arises from the second order action (\ref{action1}) for one simplex, which also shows that ${\partial \theta^\sigma_t}/{\partial A_{t'}}$ is  a symmetric matrix. Note that we obtain the same expression for the Hessian in the flat and in the homogeneously curved cases. The difficulty for the flat case is that the functions $A_t (\theta^\sigma_{t'})$ are ill-defined---only $\theta^\sigma_{t}(A_t)$ can be expected to exist locally in configuration space. However, there is no explicit expression available for these functions, and thus the route  to obtain expressions for  (\ref{HessFlat}) is by inverting $H_{\tilde t \tilde t'}^\sigma$, the matrix of (double) derivatives of the angle Gram matrix. 

From a computational perspective calculating the first and second derivatives of the determinant of the angle Gram matrix is straightforward (see Appendix \ref{SimplexK}) and the matrix $H^\sigma$ can be inverted  numerically on explicit backgrounds. The alternative  of computing $\partial \theta /\partial l$ and multiplying with the (numerically obtained) inverse of $\partial A /\partial l$  is cumbersome; although explicit expressions for $\partial \theta /\partial l$ are available \cite{DittFreidSpez}, they are quite involved.

\subsection{Identifying metric and non-metric perturbations}\label{NonGeom}

Generically, a three- or four-dimensional triangulation will have more triangles than edges and therefore we have more area variables than length variables. A single four-simplex has the same number, namely 10, of length and area variables and given the areas we can---modulo ambiguities arising from right angle configurations---compute its length variables. But, already for the case of two glued four-simplices we have 16 area and 14 length variables. Computing the lengths for each of the two simplices one will find that the lengths of the edges of the shared tetrahedron do not necessarily agree. We will refer to configurations with such a mismatch of length variables non-metric.

For the linearized theory, and for a general triangulation, we consider the matrix of derivatives
\ba\label{Gamma}
{\Gamma^t}_e:=\frac{\partial A_t }{\partial L_e} \q, 
\ea
where $A_t$ are the areas and $L_e$ are the length variables. We assume a metric background and thus the (background) length variables $L_e$ are well defined. We will use $\Gamma$ to identify the vector space of non-metric area perturbations as the space spanned by its left null vectors. Note that we can apply this characterization to all lengths and areas of a given complex, or to only the set of boundary areas and boundary lengths. In the latter case we will speak of boundary non-metricity.

\begin{figure}[t] 
   \centering
   \includegraphics[width=.6\textwidth]{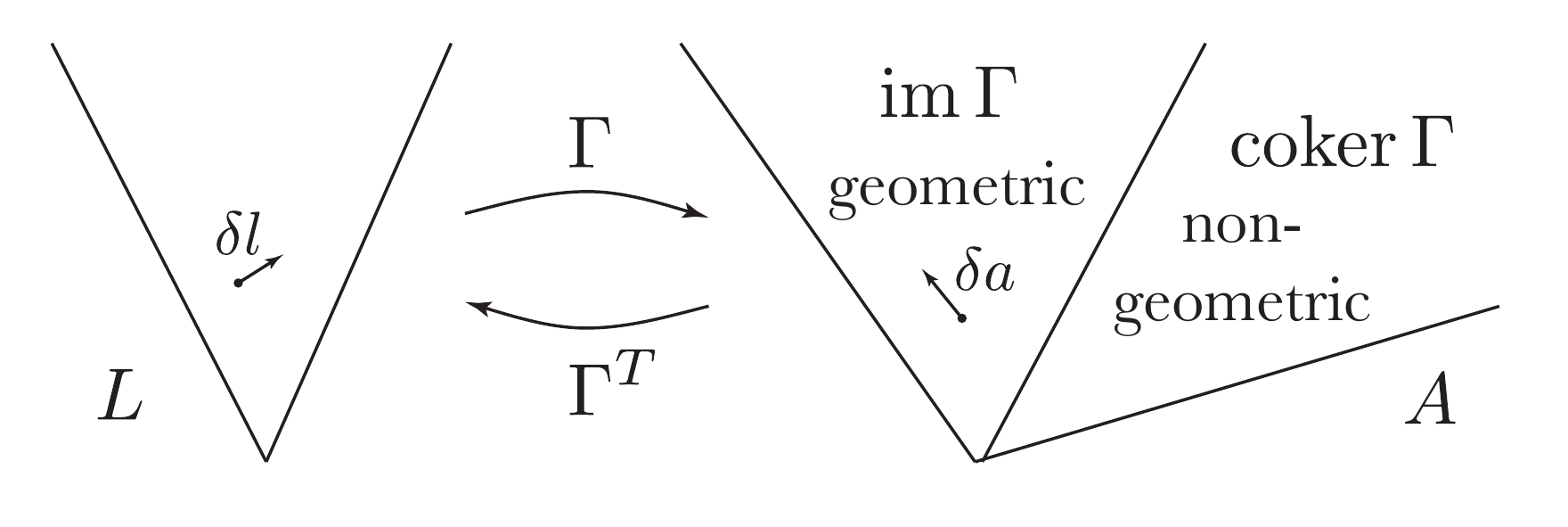} 
   \caption{The space of length perturbations maps under $\Gamma$ to the space of metric area perturbations. The complement defines the non-metric perturbations.}
   \label{fig:example}
\end{figure}

In more detail, if we call the space of edge lengths $L$ and that of the areas $A$ then $\Gamma$ can be seen as a linear map $\Gamma: TL \rightarrow TA$; these spaces are illustrated in Fig. \ref{fig:example}. In addition to the usual kernel of this map, which is spanned by the right null vectors of the matrix ${\Gamma^t}_e$,  we will also consider the cokernel, namely the quotient space $TA/\textrm{im}(\Gamma)$. Equivalently, this cokernel can be characterized as the kernel of the transpose map $\Gamma^T: TA \rightarrow TL$. Thus it is spanned by the left null vectors of ${\Gamma^t}_e$.
The cokernel can be thought of as the extra combinations of areas that go beyond the edge lengths in a given simplicial complex. More precisely, it measures the degree to which $\Gamma$ fails to be surjective. 

The dimensions of the kernel and cokernel of a linear map are not independent, 
\ba
\dim \text{coker}(\Gamma) = \dim TA - \dim \im(\Gamma) = \dim TA - \dim TL+\dim \ker(\Gamma),
\ea
which serves as a useful sanity check when you find the various null spaces.

\subsection{Gauge Symmetries}\label{gauge}

We turn to potential  gauge symmetries of the  linearized area Regge action. A quadratic action features  gauge symmetries if its Hessian has null modes that can be localized to the bulk degrees of freedom.  The presence of these null modes means that the solution under consideration is not uniquely determined by the boundary data---a gauge choice is required to uniquely specify the solution.

As emphasized in \cite{DittDiffReview}  the choice of background on which the Hessian is evaluated is important.  The number of null modes, and therefore the number of gauge symmetries, might depend on the solution being considered. In this Section we consider metric (background) solutions, while in Section \ref{diffbreaking}  we consider more general backgrounds. On the latter backgrounds we show that the gauge symmetries of the metric backgrounds are broken.

Because of the equations of motion the deficit angles $\epsilon_t$ vanish on all bulk triangles. Thus the background is given by a flat piecewise linear geometry. For any vertex $v$ in the bulk we can translate its position in the embedding flat geometry without changing the fact that the geometry is flat; in our four-dimensional triangulation there are four possible directions in which to do this. This will affect the  lengths of the adjacent edges $l_e \rightarrow l_e +\delta^I_v l_e$, with $I=1,\ldots 4$.  These translations maintain zero deficit angles and leave the boundary areas and dihedral angles invariant, that is, they do not change the intrinsic or extrinsic geometry of the boundary. Thus the area Regge action remains invariant under (bulk) vertex translations. This gives four gauge symmetries per bulk vertex.\footnote{Redundancies could occur, but only globally and thus depending on the topology of the manifold. For the four-sphere there are 10 such symmetries, 6 rotations and 4 translations. }  Note that the same kind of argument can be made if we use simplices with homogeneous curvature and the appropriate action \eqref{actioncurved}.

In our investigations of Hessians on various metric backgrounds we did not find any additional gauge symmetries. Examples can be found in section \ref{Pachnermoves}.  As the equations of motion impose flatness and would seem to suggest a topological theory, which would require more gauge symmetries, one could ask why there are not more gauge symmetries. In fact, the action
\ba\label{B-F action}
S_{\text{BF}}\,=\, \sum_t   {\cal A}_t  \, \epsilon_t( L_e)  \q ,
\ea
constructed by Baratin and Freidel \cite{B-F4D}, is topological. Here the $L_e$ are lengths associated to the edges of a triangulation, and $\epsilon_t(L_e)$ is the deficit angle calculated from these lengths. The ${\cal A}_t$ are not areas \textit{a priori}, but are treated as independent variables. Thus the  ${\cal A}_t$ are Lagrange multipliers imposing the vanishing of the deficit angles.  The equations of motion arising from variations of the length variables are
\ba\label{BFEOM}
\sum_\sigma \sum_{t\in \sigma}  {\cal A}_t  \frac{\partial \theta^\sigma_t}{\partial L_e}\,=\,0  \q .
\ea
One class of solutions is provided by the Schl\"afli identity (\ref{schlafli}): choosing ${\cal A}_t=\alpha A_t(L_e)$, where $\alpha$ is a constant, ensures that Eq. (\ref{BFEOM}) is satisfied. These solutions are called Regge solutions in \cite{B-F4D}. 

Baratin and Freidel analyze the symmetries of this action on the Regge backgrounds.  Apart from the vertex translation symmetry discussed above, there are also three symmetries per edge. These arise from perturbations of the Lagrange multipliers  ${\cal A}_t \rightarrow {\cal A}_t + {\varepsilon} n^{e',I}_t$, with $I=1,2,3$. The $n^{e',I}_t$ are specific perturbations  satisfying
\ba\label{facesymm}
\sum_t n_t^{e',I} \frac{ \partial \epsilon_t }{\partial L_e} =0   \quad \text{for all} \,\, e,
\ea
and therefore constitute an additional three gauge symmetries per edge. 
There are local redundancies between the gauge parameters, which are thoroughly discussed in \cite{B-F4D}.

Are there similar symmetries for area Regge calculus?  It turns out there are not. Here we have to consider the Hessian
$ \partial \epsilon_t/ {\partial A_{t'}}$.
This Hessian can be obtained from 
\ba\label{HessS}
\frac{ \partial \epsilon_t }{ \partial L_e }\,= \, \sum_\sigma   \frac{ \partial \theta^\sigma_t }{ \partial L^\sigma_e }
\ea
 by multiplying the Hessians associated to each simplex  by $\partial L_e^{\sigma} /\partial A_t$.  However, for the action (\ref{B-F action}), the condition $L^\sigma_e=L_e$ is imposed on all edges, which is not the case for the area Regge action. Instead when we multiply these Hessians by $\partial L_e^{\sigma} /\partial A_t$ there are more equations to satisfy, precisely one for each triangle. 
 
 Indeed in the numerical examples studied in the next section we find that ${ \partial \epsilon_t }/{ \partial L_e }$ has the (left)  null vectors resulting from Eq. (\ref{facesymm})  whereas  ${\partial \epsilon_t}/ {\partial A_{t'}}$ has only the  null vectors (which are null from the left and right) resulting from the vertex translation symmetry. The disappearance of the left null vectors in going from ${\partial \epsilon_t}/{\partial L_e}$ to ${\partial \epsilon_t}/{\partial A_{t'}}$ is a consequence of the fact that there are more areas than length variables and thus more conditions to satisfy in order to be a left null vector.

\section{Pachner moves}\label{Pachnermoves}

In this section we consider certain aspects of the dynamics of linearized area Regge calculus. Specifically, we will study the equations of motion on small simplicial complexes, those that support minimal Pachner moves.  Pachner moves are local changes of the bulk triangulation that finitely generate any change of the bulk triangulation, \cite{Pachner}. 

 In $d$ dimensions there are $(d+1)$ different types of moves referred to as  $(d+1) - 1 , d-2, \cdots ,1-(d+1)$. An $x-y$ Pachner move with $x+y = d+2$ changes a complex of $x$  $d$--simplices into a complex of $y$  $d$--simplices. A $y-x$ move is the inverse of an $x-y$ move. In four dimensions we have therefore the $5-1$ move, the $4-2$ move, their inverses, and the $3-3$ move. The subsections below treat each of these moves in turn. 
 
 The Pachner moves allow us to efficiently check the symmetry content of the theory. We also compare the behavior of the area and length Regge calculi under Pachner moves and find that there are significant differences. This could be a useful test to classify spin foam models. 
 
 Note that for this analysis we only consider the linearized theory on flat   and {\it metric} background solutions. Explicit expressions for the Hessians we have used can be found in our open-source Area Regge Calculus (ARC) Mathematica code, available at \url{https://github.com/Seth-Kurankyi/Area-Regge-Calculus}.

\subsection{5--1 move}

\begin{figure}[htbp]
\begin{tikzpicture}[scale = 2]
    \draw[thick] (0,1)--(-0.95,0.31)--(-0.59,-0.81)--(0.59,-0.81)--(0.95,0.31)--cycle;
    \draw[thick]   (-0.95,0.31)--(0.95,0.31)--(-0.59,-0.81)--(0,1)--(0.59,-0.81)--cycle;
    \node [above] at (0,1) {$1$};
    \node [right] at (0.95,0.31) {$2$};
    \node [below right] at (0.59,-0.81) {$3$};
    \node [below left] at (-0.59,-0.81) {$4$};
    \node [left] at (-0.95,0.31) {$5$}; 
    
   \draw[->,ultra thick] (2.2,-0.2)--(1.6,-0.2);
   \draw[->,ultra thick] (1.6,0.2)--(2.2,0.2);
    
    \node[above] at (1.9,0.2) {1--5};
     \node[above] at (1.9,-0.2) {5--1};
     \begin{scope}[xshift=4cm]
    \draw[thick] (0,1)--(-0.95,0.31)--(-0.59,-0.81)--(0.59,-0.81)--(0.95,0.31)--cycle;
    \draw[thick]   (-0.95,0.31)--(0.95,0.31)--(-0.59,-0.81)--(0,1)--(0.59,-0.81)--cycle;
    \draw[red,dashed,thick]  (0,1)--(0,0)--(-0.95,0.31) (-0.59,-0.81)--(0,0)--(0.59,-0.81) (0.95,0.31)--(0,0);
    \node [right] at (0.05,-0.045) {$0$};
    \node [above] at (0,1) {$1$};
    \node [right] at (0.95,0.31) {$2$};
    \node [below right] at (0.59,-0.81) {$3$};
    \node [below left] at (-0.59,-0.81) {$4$};
    \node [left] at (-0.95,0.31) {$5$}; 
    \end{scope}
\end{tikzpicture}
\caption{A 1--5 move splits a 4-simplex into five 4-simplices by introducing a bulk vertex $0$. The 5--1 Pachner move is the inverse and reduces the five 4-simplices on the right to the one 4-simplex at left by removing the bulk vertex and its associated bulk edges (dashed).}\label{5-1move}
\end{figure}
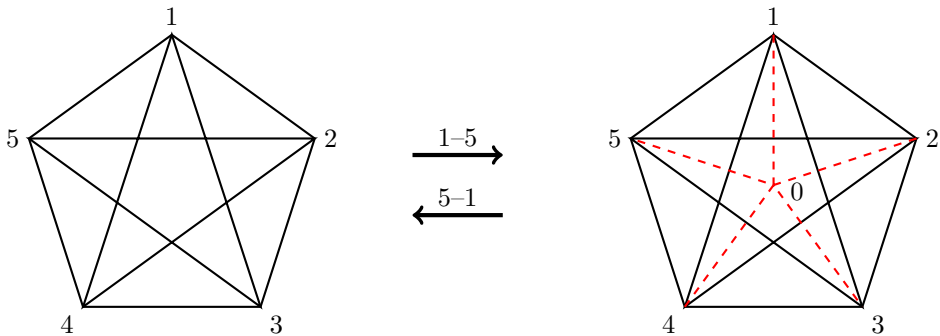

We start with the 5--1 move. Its properties in area Regge calculus turn out to be very similar to the 5--1 move in length Regge calculus. The 5--1 move starts with an initial configuration of 5 four-simplices sharing a common vertex in the bulk of the triangulation and removes the bulk vertex to get a single simplex (see Fig. \ref{5-1move}). As in length Regge calculus, a solution with 5 four-simplices is most simply constructed by subdividing a single flat simplex.  There is a free choice in this subdivision, namely where to place the bulk vertex. This leads to the four-dimensional gauge symmetry discussed above and can be seen as a remnant of the continuum diffeomorphisms. 

The simplicial complex for the initial configuration of the 5--1 move has 20 triangles and 15 edges. Of these, 10 triangles and 10 edges are in the boundary, which coincides with the boundary of a four--simplex. Thus we have 5 bulk edges and 10 bulk triangles. 

We label the vertices of our background solution by $(0,1,\ldots,5)$, with $0$ the bulk vertex. The edge lengths are chosen $l_{\text{bdry}}=1$ for edges $(i,j)$ and $l_{\text{bulk}}=\sqrt{2/5}$ for edges $(0,i)$ with $i,j \in \{1,\ldots,5\}$.  
The effective Hessian matrix describing the linearized action in terms of area perturbations is
\ba
M^{51}_{tt'}:=-\frac{1}{2}\sum_\sigma ((H^\sigma)^{-1})_{tt'},
\ea
 as described in section \ref{LinSim}, and can be found explicitly for our background.\footnote{This, and all other explicit calculations can be found in our open-source \href{https://github.com/Seth-Kurankyi/Area-Regge-Calculus}{Area Regge Calculus (ARC)} code.}
 
As expected, see the discussion in Section \ref{gauge}, the bulk part of the Hessian has exactly four null vectors; these correspond to the four vertex translations of the bulk vertex and are the discrete remnant of  diffeomorphism symmetry. The full Hessian has five null vectors: the four null vectors describing vertex translations and a global scaling symmetry, that affects also the boundary areas. 

We have also considered the (linearized) theory with homogeneously curved simplices. In this case one also finds four null vectors for the bulk Hessian. There is however no global null vector, as the scaling symmetry is broken by the homogeneous curvature. Again see the \href{https://github.com/Seth-Kurankyi/Area-Regge-Calculus}{ARC code} for the explicit computations in both the flat and homogeneously curved cases.

\subsubsection{Metricity}

As explained in Section \ref{NonGeom}, the area perturbations split into metric and non-metric types. For the 5--1 move all boundary area perturbations are metric (away from right angle configurations). Conversely, the boundary length perturbations determine the boundary area perturbations uniquely. Considering the full complex, including bulk areas, there are 5 non-metric area perturbations.

On our chosen background the solutions to the equations of motion are orthogonal to these non-metric directions. In fact, this holds for general 5--1 backgrounds due to the subdivision construction introduced above: The boundary data specify (away from right angle configurations) a metric 4-simplex and thus do not admit non-metric directions. Meanwhile, the equations of motion impose flatness for the deficit angles appearing in a subdivision of this metric simplex. Thus the subdivision determines a 4-parameter set of flat and metric solutions.

This argument shows that, like the length Regge action, the area Regge action is invariant under the 5--1 Pachner move. In particular, evaluating the action for a complex consisting of five simplices on the solution for the bulk variables we find the same value as for the final configuration consisting of only one simplex.

%
%

\subsection{4--2 move}\label{42move}

The 4--2 move starts with a configuration of four 4-simplices sharing a common edge and, by removing the common edge, ends up with a configuration of two 4-simplices glued along a (new) shared tetrahedron. 

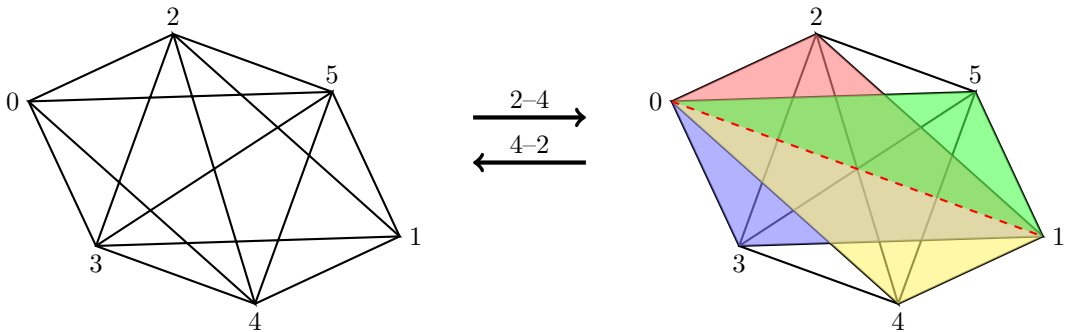
\begin{figure}[htpb]
\begin{tikzpicture}[scale = 1.5]
    \begin{scope}[rotate = -20]
    \draw[thick] (0,1)--(-1,0)--(0,-1)--(1.5,-1)--(2.5,0)--(1.5,1)--cycle;
    \draw[thick]   (1.5,1)--(-1,0)--(1.5,-1)--(0,1)--(2.5,0)--(0,-1)--(0,1) (1.5,-1)--(1.5,1)--(0,-1);
    \node [left] at (-1,0) {$0$};
    \node [above] at (0,1) {$2$};
    \node [below] at (0,-1) {$3$};
    \node [below] at (1.5,-1) {$4$};
    \node [above] at (1.5,1) {$5$};
    \node [right] at (2.5,0) {$1$};
     \end{scope}
    \draw[->,ultra thick] (4,-0.2)--(3,-0.2);
   \draw[->,ultra thick] (3,0.2)--(4,0.2);
    \node[above] at (3.5,0.2) {2--4};
     \node[above] at (3.5,-0.2) {4--2};
     \begin{scope}[xshift=5.7cm,rotate = -20]
    \draw[thick] (0,1)--(-1,0)--(0,-1)--(1.5,-1)--(2.5,0)--(1.5,1)--cycle;
    \draw[thick]   (1.5,1)--(-1,0)--(1.5,-1)--(0,1)--(2.5,0)--(0,-1)--(0,1) (1.5,-1)--(1.5,1)--(0,-1);
   \path[fill= red!50,opacity= 0.6] (-1,0)--(2.5,0)--(0,1);
    \path[fill= blue!50,opacity= 0.6] (-1,0)--(2.5,0)--(0,-1);
    \path[fill= green!70,opacity= 0.6] (-1,0)--(2.5,0)--(1.5,1);
    \path[fill= yellow!70,opacity= 0.6] (-1,0)--(2.5,0)--(1.5,-1);
    \draw[red, thick,dashed] (-1,0)--(2.5,0);    
    \node [left] at (-1,0) {$0$};
    \node [above] at (0,1) {$2$};
    \node [below] at (0,-1) {$3$};
    \node [below] at (1.5,-1) {$4$};
    \node [above] at (1.5,1) {$5$};
    \node [right] at (2.5,0) {$1$};
    \end{scope}
\end{tikzpicture}
\caption{ The 2--4 Pachner move takes two 4-simplices ${\sigma}^0=(1,2,3,4,5)$ and ${\sigma}^1 =(0,2,3,4,5)$, which share the boundary tetrahedron $(2,3,4,5)$, to four simplices ${\sigma}^2,{\sigma}^3,{\sigma}^4,{\sigma}^5$ by introducing a bulk edge $e(0 1)$. The inverse procedure gives the 4--2 move.  }\label{4-2move}
\end{figure}

The 4--2 move in area Regge calculus has quite different properties from that in length Regge calculus. The main reason is that the boundary of the complex, which agrees with the boundary of two glued 4-simplices, admits non-metric data. As we will see this possibility will be responsible for the non-invariance of the area Regge action under the 4--2 move. The boundary data for the 4--2 move in length Regge calculus, which are given by the lengths of the edges of the two glued simplices, do not induce curvature. This means that the solution for the initial configuration of the 4--2 move is  flat, and leads to the invariance of the length Regge action.

The simplicial complex for the initial configuration of the 4--2 move  has 20 triangles and 15 edges. Of these, 16 triangles and 14 edges are in the boundary, which coincides with the boundary of two glued four--simplices. Thus we have one bulk edge and four bulk triangles (see Fig. \ref{4-2move}).

Again we label vertices $(0,1,\ldots,5)$ with $(01)$ the bulk edge. The edge lengths are $l_{\text{bdry}}=1$ for edges $(i,j)$ and edges $(I,i)$ with  $i,j \in \{2,\ldots,5\}$ and $I\in\{0,1\}$.  For the bulk edge $(0,1)$, $l_{\text{bulk}}=\sqrt{5/2}$.

The solution for the bulk areas is unique, that is, the bulk Hessian has no null vectors. The full Hessian has one null vector corresponding to a global scaling symmetry.

\subsubsection{Metricity}

Amongst the 20 area perturbations of the full complex, five combinations describe non-metric directions. These are determined by the left null vectors of the matrix ${\Gamma^t}_e$. When restricted to the boundary triangles and edges, there are two null vectors and hence two non-metric directions. In both cases there are no right null vectors, which means the metric area perturbations determine the length perturbations uniquely.

Let us first consider a restriction to metric boundary perturbations. (We remind the reader that these are over a metric background.) We again find a solution by subdivision in a flat embedding. The embedding determines the bulk edge of the subdivision and as before the action is invariant when restricted to metric boundary data. 


However, this changes if we consider non-metric boundary data.  These can be isolated through a variable transformation for the boundary variables; the new variables will also be useful for the analysis of the 4--valent tent move in section \ref{Tentmoves}. The 16 boundary areas can be taken to define the areas of two simplices $\sigma^0=(0,2,3,4,5)$ and $\sigma^1=(1,2,3,4,5)$ that share a tetrahedron $\tau=(2,3,4,5)$.  This allows us to consider the following transformation
\ba
(\{a_{0ij}\},\{a_{1ij}\}, \{a_{kij}\}) \,\, \rightarrow \,\, (\{ l_{0i}\} , \{ \phi^0_\alpha\}, \{l_{1i}\},  \{ \phi^1_\alpha\},  \{a_{kij}\} )\q,
\ea
where the indices $i,j,k$ take values in $\{2,3,4,5\}$. The variables $\phi^0_\alpha$  with $\alpha \in \{1,2\}$ are two 3D dihedral angles at non-opposite edges in the tetrahedron $\tau$ and are determined by the areas of the simplex $\sigma^0$. The $\phi^1_\alpha$ describe dihedral angles at the same edges, but are computed from the areas of the simplex $\sigma^1$. Finally the $l_{mn}$ are the edge lengths between vertices $m$ and $n$. We can construct such a transformation by splitting it into two steps: For each simplex we first transform the 10 areas to the 10 length variables. We then consider separately the 6 length variables $l_{ij}$, which determine the tetrahedron $\tau$ in $\sigma^0$ and in $\sigma^1$. From these 6 length variables we can define the four areas $a_{kij}$ and the  dihedral angles $\phi^{0}_\alpha$ and $\phi^{1}_\alpha$.  See Appendix \ref{lengthTOareasangles} for the explicit transformation needed in the second step.

Clearly, we have a non-metric configuration if $\phi^0_\alpha \neq \phi^1_\alpha$, as these dihedral angles describe the same geometric quantity, but are computed from the data of different 4-simplices.  The metricity condition is thus that these 3D dihedral angles in the shared tetrahedron coincide. This is analogous to the metricity condition for tetrahedra  identified in \cite{AreaAngle}, which demands that the 2D dihedral angles of shared triangles should coincide. 

We can now introduce two variables $t_\alpha=\phi^1_\alpha-\phi^0_\alpha$ which isolate the non-metric directions.  For boundary data that give non-vanishing $t_\alpha$ the action will fail to be invariant under the 4--2 move.  Interestingly, the effective action for the configuration with four simplices couples the angles $\phi^1_\alpha$ and $\phi^0_\alpha$. This coupling cannot appear for the action with two simplices, as it is just a sum of two terms $S(\sigma^0)$ and $S(\sigma^1)$ which only depend on the quantities in $\sigma^0$ and $\sigma^1$ respectively. 

In summary, for general boundary data the area Regge action is not invariant under the 4--2 Pachner moves and the reason is that the boundary admits non-metric perturbations.

\subsection{3--3 move}

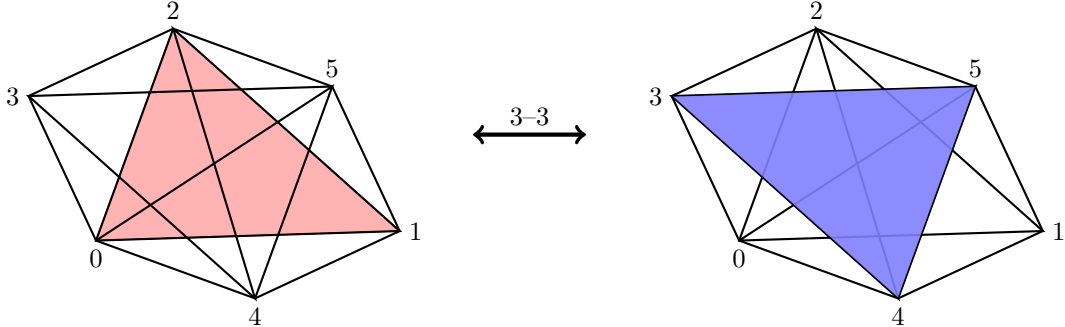
\begin{figure}[htpb]
\begin{tikzpicture}[scale = 1.5]
    \begin{scope}[rotate = -20]
    \draw[thick] (0,1)--(-1,0)--(0,-1)--(1.5,-1)--(2.5,0)--(1.5,1)--cycle;
    \draw[fill = red!30] (0,1)--(0,-1)--(2.5,0)--cycle;
    \draw[thick]   (1.5,1)--(-1,0)--(1.5,-1)--(0,1)--(2.5,0)--(0,-1)--(0,1) (1.5,-1)--(1.5,1)--(0,-1);
    \node [left] at (-1,0) {$3$};
    \node [above] at (0,1) {$2$};
    \node [below] at (0,-1) {$0$};
    \node [below] at (1.5,-1) {$4$};
    \node [above] at (1.5,1) {$5$};
    \node [right] at (2.5,0) {$1$};
     \end{scope}
    \draw[<->,ultra thick] (4,0)--(3,0);
    \node[above] at (3.5,0) {3--3};
     \begin{scope}[xshift=5.7cm,rotate = -20]
    \draw[thick] (0,1)--(-1,0)--(0,-1)--(1.5,-1)--(2.5,0)--(1.5,1)--cycle;
    \draw[thick]   (1.5,1)--(-1,0)--(1.5,-1)--(0,1)--(2.5,0)--(0,-1)--(0,1) (1.5,-1)--(1.5,1)--(0,-1);
    \draw[fill = blue!50,opacity = 0.9]  (1.5,1)--(-1,0)--(1.5,-1)--cycle;
    \node [left] at (-1,0) {$3$};
    \node [above] at (0,1) {$2$};
    \node [below] at (0,-1) {$0$};
    \node [below] at (1.5,-1) {$4$};
    \node [above] at (1.5,1) {$5$};
    \node [right] at (2.5,0) {$1$};
    \end{scope}
\end{tikzpicture}
\caption{ The 3--3 Pachner move changes a configuration of three simplices with one bulk triangle $t(0,1,2)$ to a different configuration of three simplices with a bulk triangle $t(3,4,5)$ keeping the boundary geometry fixed.  }\label{3-3move}
\end{figure}

The 3--3 Pachner move  transforms a configuration of three 4-simplices in a triangulation to a different configuration also made up of three simplices. See Fig. \ref{3-3move}. 

In length Regge calculus the 3--3 move is the only one that admits curvature. The boundary lengths can be chosen such that the bulk triangle has a non-zero deficit angle. Note that there is no bulk edge and hence no equation of motion to impose in length Regge calculus. The presence of curvature leads to non-invariance of the action under 3--3 Pachner moves \cite{DittKaminStein}.

This leads to a puzzling question for area Regge calculus: if we can choose boundary data that lead to curvature in the bulk, how are the equations of motion, which demand flatness, imposed? The resolution will be a subtle interplay between the geometric data described by the boundary areas on the one hand and the geometric data described by the boundary lengths on the other.

There are {\it only} 19 triangles in this complex---the triangle $(3,4,5)$ does {\it not} appear. There are 15 edges and the boundary includes all 15 edges and 18 of the  triangles. There is one bulk triangle.

We consider again the vertices $(0,1,\ldots,5)$ and the three four-simplices $\sigma(01234),\sigma(01235)$ and $\sigma(01245)$  which share the bulk triangle  $t(012)$.  For our background solution the edge length are $l_{\text{bdry}}=1$ for edges $(i,j)$ and $(I,J)$ with $i,j\in \{3,\ldots,5\}$ and  $I,J \in \{0,1,2\}$. The edges   $(I,i)$ have length $l_{\text{bdry}}=\sqrt{2/3}$. 

With our background solution we  find that the solution for the bulk area is unique and that the full Hessian has one null vector corresponding to the global scaling symmetry.

\subsubsection{Metricity}

As before the surfeit of area variables allows for area perturbations that are non-metric. These behave much as before and rather than treat them in detail we will focus on an issue that is unique to the 3--3 move. 

We have just shown that the boundary data uniquely determine the bulk triangle. Thus even for a metric boundary perturbation of the areas that one would expect to lead to boundary edge lengths that induce curvature we find a particular bulk triangle. As all area solutions require flatness, it seems the only possibility is that this solution is non-metric.


However, it turns out that this is not the case!

Considering ${\Gamma^t}_e$ we have now to take into account one triangle less than for the other complexes.  It has a co-image of dimension four, but, none of these non-metric perturbations has a component in the bulk area. The co-image is orthogonal to the vector describing the solution (i.e. the bulk row of the effective Hessian).

Restricting ${\Gamma^t}_e$ to boundary triangles and boundary edges we find, surprisingly, that although it is an $18\times15$ matrix, it has a one--dimensional kernel.  This means that despite their number, the 18 boundary areas do {\it not} uniquely determine the 15 boundary edge lengths. If we add the bulk area to the boundary areas---and the resulting set is metric---it will uniquely determine the (boundary) length perturbations. 

Thus, if we solve the equations of motion for a fixed and metric  area-perturbation, we actually fully determine the corresponding (boundary) length perturbation, and, as it must, it determines a zero curvature solution. 

~\\
Considering the background solution cited above one finds that the area Regge action is invariant under the 3--3 move. That is, the actions for both configurations of the 3--3 move agree after one has included, for each configuration, the bulk area. However, this invariance is due to the highly symmetric nature of the background. We have studied less symmetric backgrounds and found that the area Regge action is not invariant. This can again be traced back to the non-metric boundary perturbations.

\subsection{Summary: Pachner moves in area vs. length Regge calculus}

In summary, we have found that both for length and area Regge calculus only the initial configuration of the 5--1 move features gauge symmetries; these are remnants of the diffeomorphism symmetry.  Both actions are also invariant under the 5--1 move. 

For the other Pachner moves one finds however differences:  the boundary configuration for the 4--2 Pachner moves admits non-metric directions. This leads to a non--invariance of the area Regge action. In contrast, the length Regge action is invariant under the 4--2 Pachner move as the boundary configuration admits only flat solutions.

The 3--3 move is the only one under which the length Regge action is not invariant. The reason is that the boundary length data generically prescribe a non-vanishing deficit angle for the bulk triangle. A vanishing deficit angle requires special boundary data. There is no bulk edge and thus no equation to solve in length Regge calculus.

There is however a bulk triangle and thus an equation of motion---imposing flatness---in area Regge calculus. Given the fact that the length boundary data generically induce curvature one might wonder how this flatness is realized. It turns out however that the area boundary data do not fully determine the length boundary data even though there are 18 areas and only 15 lengths in the boundary. In fact, it is exactly the bulk area that is needed to fully determine the boundary lengths, and its value is fixed by the equation of motion which demands a vanishing deficit angle. 

Nevertheless, the area Regge action is also (generically) not invariant under the 3--3 move. The reason again is that non-metric area perturbations can appear in the area boundary data.

\section{Tent moves} \label{Tentmoves}

We now analyze the linearized dynamics of area Regge calculus in a canonical framework. This will tell us whether, apart from the constraints expected from the vertex translation symmetries, there are any further (e.g. second class) constraints, that would reduce the number of physical degrees of freedom. We will find that this is not the case, and that, as the number of area variables is typically larger than the number of length variables, area Regge calculus has a larger number of physical degrees of freedom than length Regge calculus.

The canonical framework we will be employing, \cite{DittrichHoehn1}, uses discrete time evolution steps. The action will serve as a generating function for the canonical transformation that represents the time evolution \cite{GambiniPullin}. This has the advantage that the covariant equations of motion are exactly reflected in the canonical framework. The symmetry content is also mirrored exactly \cite{DittrichHoehn1}, and thus we will find that the vertex translation symmetry leads to (first order) constraints. The latter can be used to define a notion of continuous time transformations. (Again here we consider metric background solutions and these features will only hold on such backgrounds.)

Having chosen a discrete evolution we need to decide how to evolve the triangulation stepwise. We use local evolution moves, which do not change the connectivity of the spatial triangulation, and are called tent moves \cite{tentmoves}.  Consider all tetrahedra $\{\tau_i\}$  that share the vertex $v_0$ in the triangulation of an equal time hypersurface. The union of these tetrahedra defines the three-dimensional star of $v_0$. We glue an edge $e(v_0,v_1)$---the tent pole---to $v_0$ and thus obtain a new vertex $v_1$.  This new vertex will be connected by new edges with all vertices adjacent to $v_0$ in the initial (equal-time) triangulation.   For each $\tau_i$ we glue a simplex $\sigma_i$ onto $\tau_i$ so that all these simplices share the tent pole $e$.  The tent moves change the geometric data of the triangulated hypersurface by replacing the three-dimensional star of $v_0$ with the three-dimensional star of $v_1$.  

The canonical framework developed in \cite{DittrichHoehn1} provides a setting in which to analyze these tent moves.\footnote{The more general framework of \cite{DittrichHoehn1} also allows for Pachner moves of the spatial triangulation. These can be seen as time evolution steps in which the number of degrees of freedom change.} To this end one needs to consider the action $S_{\rm T}$ associated to the triangulation piece  ${\rm T}$ that is glued onto the hypersurface during the tent move.  This piece of triangulation carries 4 types of variables: 
\begin{enumerate}
\item[(a)] Variables associated to the `lower' boundary of ${\rm T}$ that will be glued  and thus `disappear'. These variables are associated to the initial time and the simplices sharing the vertex $v_0$.
\item[(b)] Exactly the same number of `new' variables are associated to the `upper' boundary of ${\rm T}$, and can be associated to the final time.
\item[(c)] There will also be variables associated to the corner of the tent ${\rm T}$; these are variables that appear at both the initial and final time. They do not change under the tent move evolution and, hence, are  non-dynamical.
\item[(d)] Finally, there are variables associated to the bulk of the tent ${\rm T}$. In length Regge calculus the only such variable is the length of the tent pole, while in area Regge calculus there are all the areas of the triangles that hinge on the tent pole. 
\end{enumerate}

The bulk variables can be incorporated into the canonical framework, but the conjugated momenta will always be constrained to vanish. This imposes the equations of motion for the bulk variables, and coincides with the covariant equations of motion.  Using the solutions in the remaining equations gives a reduced phase space. Equivalently one can integrate out the bulk variables from the action. This leaves only the variables of the types (a), (b) and (c). We will proceed along the latter path and denote by $S_{\rm T}$ the (effective) action with the bulk variables integrated out. 

The `corner' variables, type (c), are non-dynamical and do not have associated momenta. For the remaining variables,  $x_0^i$ at the initial time and $x_1^i$ at the final time, the equations
\ba\label{CanEqu}
p_0^{i}\,=\, -\frac{\partial S_{\rm T}}{\partial x_0^i} \q ,\q\q p_1^{i}\,=\, \frac{\partial S_{\rm T}}{\partial x_1^i}  \quad 
\ea
serve both to define the momenta and as equations of motion. Solving for the final data $(x^i_1,p^i_1)$ in terms of the initial data $(x^i_0,p^i_0)$ proceeds in two steps: the first set of equations in \eqref{CanEqu} is solved for the $x^i_1$ and then these solutions  are used in the second set to find the $p^i_1$. 

It can, however, happen that a solution of the first set of equations in terms of $x^i_1$ is not possible. Similarly one might not be able to solve the second set of equations for $x^i_0$.  This will be the case if the matrix 
\ba\label{SMixed}
\frac{\partial^2 S_{\rm T}}{\partial x_0^i \partial x_1^j}
\ea
is not invertible, i.e. we have a degenerate Lagrangian system.  The matrix will thus have left null vectors $Y^i_I$. By contracting the first set of equations in (\ref{CanEqu}) with such a null vector 
\ba\label{Constraintsgeneral}
\sum_i Y^i_I p_0^i\,=\, - \sum_i Y^i_I \frac{\partial S_{\rm T}}{\partial x_0^i}
\ea
we can project out any linear dependence on perturbations in the variables $x_1^j$ around points where \eqref{SMixed} is non-invertible. In fact, for a linear theory (\ref{Constraintsgeneral}) these equations are linear and thus they will  lead directly to constraints.\footnote{An in-depth analysis of the various types of constraints that can appear can be found in \cite{DittHoehn2}.} These constraints are equations (of motion) that hold between the configuration variables and momenta at one time.  Thus left null vectors of (\ref{SMixed}) lead to constraints for the data at the initial time.  Similarly, right null vectors $Z_I^j$ of (\ref{SMixed}) will lead to constraints at the final time.  Gauge symmetries, which correspond to localizable null vectors for the (bulk) Hessian of the action, always lead to constraints \cite{DittrichHoehn1}. The gauge constraints are preserved by the time evolution, or more precisely, initial data satisfying the initial constraints will be mapped to data satisfying the final constraints. Correspondingly, the equations of motion will not lead to a unique solution for the final data in terms of the initial data. Instead we have a gauge freedom, involving the same number of parameters as we have constraints. This is an expected consequence of the gauge symmetry of the action. 

The main result of our tent move analysis is that we find only constraints resulting from the gauge symmetry of the action. As in the continuum, we will differentiate between gauge and physical degrees of freedom. The evolution of the physical degrees of freedom---in contrast to that of the gauge degrees of freedom---is determined by the tent move equations of motion.  More precisely these are phase space functions that Poisson commute with the constraints (with the canonical Poisson structure $\{x^i,p^j\}=\delta^{ij}$ between variables at one time). 

Note that the definition of a physical degree of freedom depends on the notion of tent move. For example, we might find that tent moves have physical degrees of freedom, whereas a more global notion of time evolution might find only gauge degrees of freedom. The reason is that we consider the `corner' variables (type (c) above)  as constant and thus freeze gauge symmetries that affect these variables (or variables outside the region of the tent move).  In contrast,  the notion of a gauge degree of freedom will remain the same even with a more global time evolution.  

As an example, consider the degrees of freedom of length Regge calculus in $(2+1)$ and $(3+1)$ dimensions. The $(2+1)$-dimensional Regge calculus is topological: using a global time evolution one finds that the number of physical degrees of freedom does not depend on the size of the triangulation, but only on the topology of the underlying space.  There are no local physical degrees of freedom.  However, a tent move over a vertex with $n$ adjacent edges, which we call an $n$-valent tent move, will have $n-3$ physical and three gauge degrees of freedom.  The appearance of physical degrees of freedom for the tent moves is, nevertheless, consistent with the finding that there are no local physical degrees of freedom under a global time evolution. The tent moves show that there are three constraints (or three gauge degrees of freedom) per vertex and modulo a topological constant this agrees with the number of edges in a two-dimensional triangulation.

Similarly, for $(3+1)$--dimensional Regge calculus an $n$--valent tent move gives $n-4$ physical and four gauge degrees of freedom. In this case, one also finds local physical degrees of freedom under global time evolution---the reason is that for a sufficiently large three-dimensional triangulation the number of edges is generically greater than four times the number of vertices.\footnote{The exceptions are so-called stacked triangulations of the three-sphere, see \cite{DittrichHoehn1}.}

In the examples considered here we find no additional constraints, beyond those resulting from the vertex translation symmetry. There are more dynamical area than dynamical length variables, and so we have more physical degrees of freedom for area Regge calculus than in the length calculus. This will also hold  in a global time evolution, as generically there are more triangles than edges in  three-dimensional triangulations. We have thus more kinematical variables in area than in length Regge calculus, but the same gauge freedoms (in the theories linearized on a metric or flat background respectively).  

Here we have studied the 4-valent and 5-valent tent moves in detail. These already exemplify all the dynamical features that appear in length Regge calculus and those that we expect to appear in the area calculus. In length Regge calculus the 4-valent tent move only admits a flat dynamics and all four dynamical degrees of freedom turn out to be gauge. In contrast, in the area calculus the 4-valent tent move has six dynamical degrees of freedom, of which four are gauge and two are physical; the latter represent non-metric degrees of freedom. 

The 5-valent tent move in length Regge calculus admits curvature and out of the five dynamical degrees of freedom one is physical. In area Regge calculus we find four gauge and five physical degrees of freedom. Four of these physical degrees of freedom describe non-metric motions. 

One can obtain the equal time triangulated hypersurfaces for higher valent tent moves from those for lower valent tent moves by subdividing tetrahedra adjacent to the vertex $v_0$ with  1--4 Pachner moves. In going from an $n$-valent to an $(n+1)$-valent tent move you add one edge and three triangles all adjacent to $v_0$. This counting shows that an $n$--valent tent move has $(3n-6)$ dynamical area variables, which can be compared to the $n$ dynamical length variables in the length calculus. As we have found no indication of gauge symmetries beyond vertex translation nor additional (possibly second class) constraints for the 4--valent and 5--valent tent moves we expect that there are only four gauge degrees of freedom for all the tent moves. This leads to $(3n-10)$ physical degrees of freedom for an $n$--valent tent move in area Regge calculus, significantly more than the $(n-4)$ physical degrees of freedom one finds in length Regge calculus.

\subsection{The 4-valent tent move}\label{4--valent move}

\begin{figure}[htpb]
\begin{tikzpicture}[scale = 1.6]
    \begin{scope}[rotate = -20]
    \draw[thick] (0,0.85)--(0,-0.85)--(1.5,-1)--(2.2,0)--(1.5,1)--cycle;
    \draw[thick]   (1.5,1)--(1.5,-1)--(0,0.85)--(2.2,0)--(0,-0.85)--(1.5,1);      
    \node [above] at (0,0.85) {$0$};
    \node [left] at (0,-0.85) {$3$};
    \node [below] at (1.5,-1) {$4$};
    \node [right] at (1.5,1) {$2$};
    \node [right] at (2.2,0) {$5$};
     \end{scope}
   \draw[->,ultra thick] (3,-0.2)--(4,-0.2);
     \node[above] at (3.5,-0.1) {$4-$valent move};
     \begin{scope}[xshift=5.7cm,rotate = -20]
     \draw[thick] (0,-0.85)--(1.3,-1)--(2.2,0)--(1.7,1)--cycle;
    \draw[thick]  (0,-0.85)--(2.2,0) (1.3,-1)--(1.7,1); 
    \draw[ thick] (0.2,0.75)--(0,-0.85) (0.2,0.75)--(1.3,-1) (0.2,0.75)--(2.2,0) (0.2,0.75)--(1.7,1);  
    \draw[blue, very thick, densely dotted] (-0.9,1.5)--(0,-0.85) (-0.9,1.5)--(1.3,-1) (-0.9,1.5)--(2.2,0) (-0.9,1.5)--(1.7,1);   
    \path[fill= red!50,opacity= 0.6] (0.2,0.75)--(-0.9,1.5)--(1.7,1);
    \path[fill= blue!50,opacity= 0.6] (0.2,0.75)--(-0.9,1.5)--(0,-0.85);
    \path[fill= green!70,opacity= 0.6] (0.2,0.75)--(-0.9,1.5)--(1.3,-1);
    \path[fill= yellow!70,opacity= 0.6] (0.2,0.75)--(-0.9,1.5)--(2.2,0);  
     \draw[red, thick,densely dashed] (0.2,0.75)--(-0.9,1.5); 
    \node [left ] at (-0.9,1.6) {$1$};
    \node [below right] at (0.2,0.75) {$0$};
    \node [left] at (0,-0.85) {$3$};
    \node [below] at (1.3,-1) {$4$};
    \node [right] at (1.7,1) {$2$};
    \node [right] at (2.2,0) {$5$};
    \end{scope}
\end{tikzpicture}
\caption{ A tent move at the 4-valent vertex 0. Introducing  a tent pole $e(01)$ and connecting the vertex 1 to the vertices in $\tau(2345)$ yields the final configuration. This introduces four bulk areas. }\label{4-Valent move}
\end{figure}
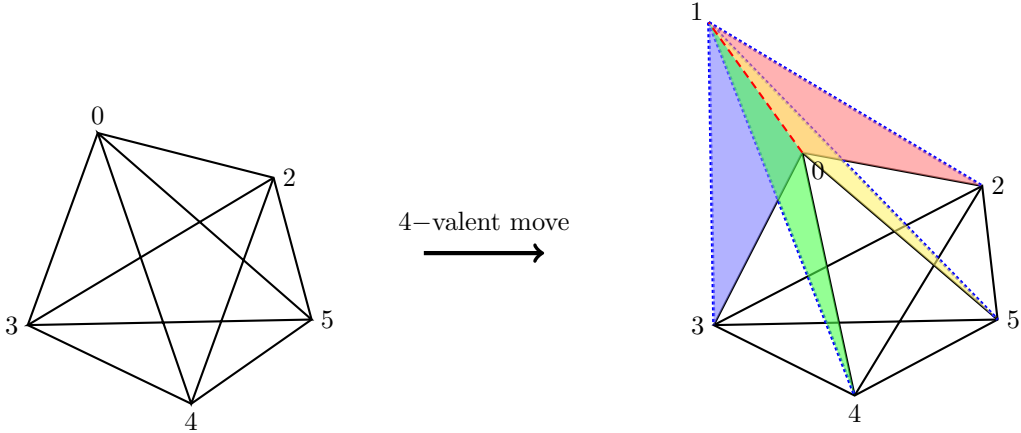

For a 4-valent tent move at a vertex $v$ we glue four 4-simplices that share the tent pole onto the 4 tetrahedra that make up the 3D star of the vertex $v$, see Fig. \ref{4-Valent move}. As these four simplices share an edge they turn out to coincide with the 4 simplex configuration of the 4--2 Pachner move.  The bulk areas that appear in the 4--2 move are the `lapse' areas of the triangles sharing the tent pole.  Thus the effective action for the 4--2 move defines also the effective action for the tent move.  We will denote this action by $S_{4V}$. Because we integrate out all bulk variables,  $S_{4V}$ only depends on the variables in the boundary of the  complex.

This relation with the 4--2 Pachner move allows us to work again with the background introduced in Section \ref{42move}. (In that background the two simplices that make up the final configuration of the 4--2 move have positive orientation. For the tent move it is more typical to consider one simplex with positive and the other (initial) simplex with negative orientation. This describes a larger tent being erected on top of a smaller base. However, the tent move is well defined for any choice of orientation. The background chosen here corresponds to putting up a tent over a pit.) 

With a tent move like that depicted in Figure \ref{4-Valent move} we have the following configuration variables:
\begin{enumerate}
\item[(a)] At time $0$ we have  six areas $a_{0ij}$  (we will use $i,j,k\in \{2,3,4,5\}$ and $\alpha\in \{1,2\}$).  In length Regge calculus we have four lengths $l_{0i}$.
\item[(b)] At time $1$ we also have six areas $a_{1ij}$ and four lengths $l_{1i}$.
\item[(c)] In addition there are four areas $a_{ijk}$ that are non-dynamical `corner' variables, and appear at both times. There would be six such variables $l_{ij}$ in length Regge calculus. 
\end{enumerate}
As for the 4--2 move, we have 16 area variables in the boundary of the tent move complex and 14 length variables. 

The boundary of the tetrahedron $\tau(2,3,4,5)$ defines the `corner' for the tent move. We therefore have to keep the four  areas of this tetrahedron constant, as its boundary defines the `corner' for the tent move.  This is different from length Regge calculus, where all six  edge lengths of this tetrahedron are fixed. Hence there are two degrees of freedom associated to this tetrahedron that are dynamical in area Regge calculus but not in the length calculus. As the areas have to be constant we can identify these two degrees of freedom as 3D dihedral angles hinging at two non-opposite edges. (Four areas and two non-opposite 3D dihedral angles determine all the lengths of a tetrahedron, see Appendix \ref{lengthTOareasangles}.) The fact that the 3D dihedral angles can change is key to non-metricity: changing an angle affects the lengths of the edges of the tetrahedron. These edges are however part of the `corner' which constitutes the boundary  of both  equal time hypersurfaces. 

We will therefore apply the same variable transformation as in the discussion for the 4--2 move, that is,  (remember $i,j,k\in \{2,3,4,5\}$ and $\alpha\in \{1,2\}$)
\ba\label{trafo2}
(\{a_{0ij}\},\{a_{1ij}\}, \{a_{ijk}\}) \,\, \rightarrow \,\, (\{ l_{0i}\} , \{ \phi^0_\alpha\}, \{l_{1i}\},  \{ \phi^1_\alpha\},  \{a_{kij}\} )
\ea
where the $(l_{0i},\phi^0_\alpha)$ and the $(l_{1i},\phi^1_\alpha)$ appear now as dynamical variables at time $0$ and time $1$, respectively. The $a_{kij}$ are the non-dynamical `corner' variables. 

The canonical evolution equations have the following form
\ba\label{42ceom}
p^l_{0i}&=&-\frac{\partial S_{4V}}{\partial l_{0i}}  \,=:\,- \sum_j M^{00}_{i,j}  l_{0j} - \sum_\alpha M^{00}_{i,\alpha} \phi^0_\alpha - \sum_{(jkl)}M^{0b}_{i,jkl} a_{jkl}  \q ,  \label{4v1} \\
p^\phi_{0\alpha}&=&-\frac{\partial S_{4V}}{\partial \phi^0_\alpha}   \,=:\,  -\sum_j   M^{00}_{\alpha,j}\,  l_{0j}   -\sum_\beta M^{00}_{\alpha,\beta}  \phi^0_\beta  -\sum_\beta M^{01}_{\alpha,\beta} \phi_\beta^1   - \sum_{(jkl)}M^{0b}_{\alpha,jkl} a_{jkl}   \q ,  \label{4v2}  \\
p^l_{1i}&=&\,\,\,\,\frac{\partial S_{4V}}{\partial l_{1i}}  \,=:\, \,\,\,\,\,\,
 \sum_j M^{11}_{i,j} \, l_{1j} + \sum_\alpha M^{11}_{i,\alpha} \phi^1_\alpha - \sum_{(jkl)}M^{1b}_{i,jkl} a_{jkl} 
\q ,  \label{4v3} \\
p^\phi_{1\alpha}&=&\,\,\,\,\frac{\partial S_{4V}}{\partial \phi^1_\alpha}   \,=:\,  \,\,\,\,\,\,
 \sum_j   M^{11}_{\alpha,j}  \, l_{1j}   +\sum_\beta M^{11}_{\alpha,\beta}  \phi^1_\beta  +\sum_\beta M^{10}_{\alpha,\beta} \phi_\beta^0   + \sum_{(jkl)}M^{1b}_{\alpha,jkl} a_{jkl} 
   \label{4v4} \q  . 
\ea

Here we made use of the fact that the mixed time block of the Hessian for the action, expressed in terms of $x^A_I$ with $A=0,1$ at time $0$ or $1$ respectively,
\ba\label{Mnd}
M^{01}_{IJ}:= \frac{\partial^2 S_{4V}}{\partial x^0_I \partial x^1_J}\q ,
\ea
 has four left and four right null vectors.  These are given by the length perturbations $l_{0i}$ and $l_{1i}$, respectively.  These null vectors result from the vertex translation symmetry discussed in Sec. \ref{gauge}. 
 
The presence of these null vectors can be explained as follows: Consider an extension of the tent move triangulation so that, e.g., the vertex $v_1$ appears as a bulk vertex of the extended triangulation.  Such an extension can be obtained via a second tent move from time 1 to time 2. The action for the extended triangulation is a sum of two terms $S_{01}$ and $S_{12}$---one for the first tent move from time 0 to time 1 and one for the second tent move between times 1 and 2. As discussed in section \ref{gauge}, the Hessian of the full action $S_{01}+S_{12}$ has four null vectors corresponding to the vertex translation symmetry of $v_1$, and thus these null vectors have entries only for variables at time 1. (We assume we have integrated out all lapse like variables.) Being null vectors for the full Hessian they are also annihilated by the non-diagonal block $M^{01}$ of the Hessian for the action $S_{01}+S_{12}$, which coincides with the non-diagonal block of the Hessian for $S_{01}$ alone. Likewise the null vectors are annihilated by the non--diagonal block $M^{21}$ of  $S_{12}$.  By time translating the argument we have that $M^{01}$ has at least four left null vectors and at least four right null vectors. 

We have  not found any further null vectors for the Hessian (\ref{Mnd}) evaluated on the background described above nor on the other backgrounds we investigated.   For the background described above $M^{00}_{i,\alpha}$ vanishes, but this is due to the high symmetry of this background and we did find non-vanishing entries on more general backgrounds.

The null vectors of $M^{01}_{IJ}$ correspond exactly to the perturbations described by the four length variables $l_{0i}$ or $l_{1j}$, as these are independent parameters for the gauge action resulting from vertex translation of   $v_0$ or of $v_1$. 

Thus equation (\ref{4v1}) only involves variables at time 0 (including the non--dynamical variables), whereas equation (\ref{4v3}) only involves variables at time 1. These constitute constraints 
\ba\label{constr4v}
C^A_i:= \,\, p^l_{Ai} + (-1)^A \sum_j M^{AA}_{i,j}  l_{Aj} +(-1)^A \sum_\alpha M^{AA}_{i,\alpha} \phi^A_\alpha  + (-1)^A\sum_{(jkl)}M^{Ab}_{i,jkl} a_{jkl}
\ea
and  these constraints are also preserved by time evolution.  The constraints (at a fixed time) are first class, i.e. they Poisson commute. This follows from the fact that the matrix $M^{AA}_{i,j}$ is symmetric.   

This also means that given a set of initial data that satisfy the constraint equations the length variables at time $1$ are not determined by the equations of motion (\ref{42ceom}). We can rather choose these freely. These four length variables represent the lapse and shift gauge degrees of freedom and describe the position of the `tip of the tent', by giving its distance to its (four) adjacent vertices.


Non-trivial dynamics will be confined to the angle variables $\phi^A_\alpha$.  Changes of these variables under time evolution means that non-metricity is being generated.  

We can however alter the dynamics and impose constraints that ensure $\phi^0_\alpha=\phi^1_\alpha$. This can be done by hand,  but a more elegant procedure is to replace the action we were considering by the action $S^2:=S(\sigma^0)+S(\sigma^1)$ for the 2 simplex configuration of the 4--2 move. (Remember that the effective actions for the 4 simplex and the 2 simplex configurations agree when projected onto the space of metric boundary perturbations.) By defining the dynamics using the action $S^2$, all variables at time step 0 decouple from the variables at time step 1. This leads to constraints for the momenta conjugated to the angle variables. These momenta at,  say, time 0 would only involve the action of the simplex $\sigma^0$.  The angle variables at time 1 will now also appear as gauge parameters and can be chosen to agree with the angle variables at time 0.

\subsection{5--valent tent move}

Next we will discuss the 5-valent tent move. In length Regge calculus this move has one physical degree of freedom and the canonical data, or equivalently, the boundary data, can be chosen so that the configuration has curvature.  As area Regge calculus imposes flatness, we expect that---as in the 3--3 move---the boundary areas do not completely fix the lengths on the boundary. 

To be more precise we consider a tent move that puts up a tent pole between $0$ and $1$. The triangulation at time $0$ can be obtained from gluing two simplices $(02345)$ and $(02346)$ along the tetrahedron $(0234)$. Note that this shared tetrahedron $(0234)$ is not part of the 3D `equal time' hypersurface. In particular, the triangle $(234)$ will not be part of the boundary data. However, the edges $(23)$, $(24)$, and $(34)$ are part of the tent move complex.

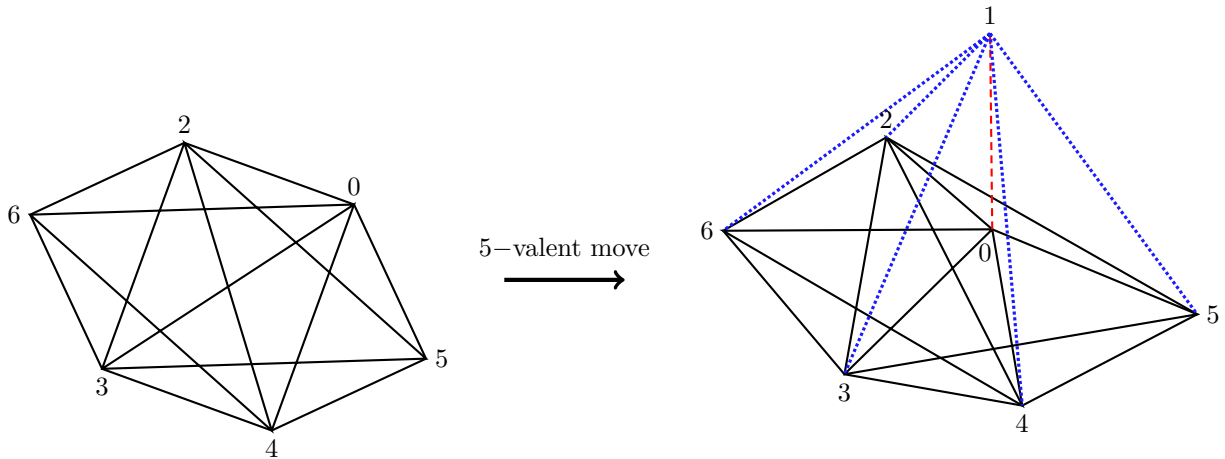
\begin{figure}[htpb]
\begin{tikzpicture}[scale = 1.6]
    \begin{scope}[rotate = -20]
    \draw[thick] (0,1)--(-1,0)--(0,-1)--(1.5,-1)--(2.5,0)--(1.5,1)--cycle;
    \draw[thick]   (1.5,1)--(-1,0)--(1.5,-1)--(0,1)--(2.5,0)--(0,-1)--(0,1) (1.5,-1)--(1.5,1)--(0,-1);
    \node [left] at (-1,0) {$6$};
    \node [above] at (0,1) {$2$};
    \node [below] at (0,-1) {$3$};
    \node [below] at (1.5,-1) {$4$};
    \node [above] at (1.5,1) {$0$};
    \node [right] at (2.5,0) {$5$};
     \end{scope}
   \draw[->,ultra thick] (3,-0.2)--(4,-0.2);
   \node[above] at (3.5,-0.1) {$5-$valent move};
     \begin{scope}[xshift=6cm,rotate = -10]
    \draw[thick] (0,1)--(-1.2,0)--(0,-1)--(1.5,-1)--(2.8,0)--(1,.4)--cycle;
    \draw[thick]   (1,.4)--(-1.2,0)--(1.5,-1)--(0,1)--(2.8,0)--(0,-1)--(0,1) (1.5,-1)--(1,.4)--(0,-1);
    \draw[red, thick,densely dashed] (1,.4)--(.7,2);
    \draw[densely dotted,very thick,blue!90]  (.7,2)--(-1.2,0) (.7,2)--(0,1) (.7,2)--(0,-1) (.7,2)--(1.5,-1)  (.7,2)--(2.8,0);
    \node [above] at (.7,2) {$1$};
    \node [left] at (-1.2,0) {$6$};
    \node [above] at (0,1) {$2$};
    \node [below] at (0,-1) {$3$};
    \node [below] at (1.5,-1) {$4$};
    \node [below] at (.95,0.35) {$0$};
    \node [right] at (2.8,0) {$5$};
    \end{scope}
\end{tikzpicture}
\caption{ A five-valent tent move at the vertex 0 starting from a configuration of two simplices $\sigma^6 = (0,2,3,4,5)$ and $\sigma^5 = (0,2,3,4,6)$ and gluing six simplices on the 3D star of vertex 0. }\label{5-Valent move}
\end{figure}

The tent move is performed by gluing 6 simplices $(01ij5)$ and $(01ij6)$, with $i,j \in \{2,3,4\}$, onto the `equal time' hypersurface.  The background we will be considering is given by $L_{0i}=L_{05}=L_{06}=L_{ij}=L_{i5}=L_{i6}=1$ and $L_{1i}=\sqrt{9/2}$ as well as $L_{15}=L_{16}=2$. The length of the tent pole can then be computed to be $L_{01}=\sqrt{3/2}$. 

In the following table we give the number of triangles for the full tent move complex and for the various sub-triangulations ($T=0$ and $T=1$ indicate the vertices $0$ and $1$ and  again $i,j \in \{2,3,4\}$):

\begin{table}[!th]
\begin{tabular}{|c|c|c|c|c|}\hline
~&\  Full complex\ \ &\ Boundary\ \ & Equal time: bulk &\ Equal time: boundary\ \ \\ \hline
Areas\,\,\, & 29 & 24&\,\, 9:\,\, $\{a_{Tij},a_{0i5}, a_{Ti6}\}$ \,\,&\,\, 6:\,\,$\{a_{ij5}, a_{ij6}\}$\,\, \\ \hline
\ Lengths\,\,\, & 20 & 19 & 5:\,\, $\{l_{Ti}, l_{T5},l_{T6}\}$ & 9:\,\, $\{l_{ij},l_{i5},l_{i6}\}$ \\ \hline
\end{tabular}
\end{table}

The boundary of the tent move complex has 24 triangles and only 19 edges.  Considering the ${\Gamma^t}_e$ matrix of derivatives of areas with respect to lengths we find one right null vector and six left null vectors. 

Restricting to the data at time $T=0$ we have $15=(9+6)$ triangles and $14=(5+9)$ edges. For this case we find that ${\Gamma^t}_e|_{T=0}$ has one right null vector and two left null vectors.  The right null vector can indeed be identified with the `missing' area $a_{234}$. That is, it represents the linearized expression for this area in terms of the length perturbations. The two left null vectors can be identified with the differences between a pair of 3D dihedral angles in the shared tetrahedron as computed from the 4-simplex containing $v=5$ and the 4-simplex containing $v=6$, respectively.

Hence, as for the 3--3 move, the area boundary data do not completely determine the (length) geometry of the boundary. This allows for a dynamics that imposes vanishing deficit angles.  

The fact that one of the areas is not available makes a transformation, similar to the one we performed for the 4-valent tent move, impossible. We already have non-metric degrees of freedom at a single time step, those picked out by the two left null vectors of ${\Gamma^t}_e(0)$ or of ${\Gamma^t}_e(1)$. Additionally there are non-metric degrees of freedom that can occur upon gluing the two time step complexes together.

As in the case of the 4-valent tent move we find exactly four null vectors for the Hessian block between variables at time $T=0$ and at time $T=1$. These four null vectors represent metric perturbations as they are in the image of the map $\Gamma$. Hence there are four constraints that result from each of the vertex translation symmetries of $v_0$ and $v_1$.  This leaves five physical degrees of freedom that split into a metric perturbation (the fifth edge length) and four non-metric perturbations. The latter describe the (two) non-metric degrees of freedom appearing within an equal time hypersurface and another two degrees of freedom describing non-metricity  due to time evolution. We also encountered this last type in the 4-valent tent move.

Length Regge calculus has only one physical degree of freedom in the 5-valent tent move: four of the edge lengths can be viewed as gauge parameters and the fifth as the physical degree of freedom.\footnote{A pair of phase space functions that commute with the constraints  and thus represent Dirac observables can be defined using the methods of \cite{DittObs}. These observables can be made to give the values of the fifth edge length and its conjugated momentum at the point in the gauge orbit where the other four edge lengths have prescribed values.} In area Regge calculus we have four additional degrees of freedom that arise from the various ways non-metricities can occur, namely, within an equal time hypersurface and as a result of time evolution.


\section{non-metricity breaks diffeomorphism symmetry}\label{diffbreaking}

For our explorations of the dynamics and of the symmetries of area Regge calculus we have so far assumed a metric background solution where the length variables can be consistently defined. For each four-simplex  $\sigma$ (and away from right angle configurations) we can define 10 functions $L^\sigma_e$, which depend on the 10 area variables $A_t$, and evaluate to the length of the edges $e$ of the simplex. Metric configurations are such that the length functions for the same edge, but coming from different four-simplices, agree.

The presence of gauge symmetries depends on the solution one is considering: the gauge symmetries specify in which ways we can deform this solution 
and still obtain a solution to the equations of motion (with the same boundary data).  

The dependence of the number of gauge symmetries on the solution appears, in particular, if we consider discretizations of continuum systems with gauge symmetries.  Often, and certainly for diffeomorphism symmetry, discretization breaks these gauge symmetries. There may be, however, special solutions, e.g. flat space in (length) Regge calculus, which exactly mirror a solution of the continuum theory. In this case the gauge symmetries around this solution are preserved.  A necessary and sufficient criterion for the existence of gauge symmetries is that the Hessian of the action, evaluated on this solution, has null vectors localized to the bulk degrees of freedom.  
Moving away from these special solutions there is no guarantee that the gauge symmetries still exist. 

For (length) Regge calculus reference \cite{WilliamsR}  identified the vertex translation symmetries  as gauge symmetries using a flat background. This work also  showed that the vertex translation symmetries can be matched in a continuum limit to the diffeomorphism symmetry of the continuum. Motivated by these findings, \cite{HamberWilliams} and other references argued that the vertex translation symmetries exist generally for (length) Regge calculus, i.e. for arbitrary backgrounds.   

This turned out not to be the case. Reference \cite{BahrDittrich1} considered solutions with curvature and explicitly evaluated the Hessian of the Regge action on these solutions. This showed that the vertex translation symmetries are broken by curvature. More precisely, in the example considered in \cite{BahrDittrich1} the lowest eigenvalues grew quadratically with a deficit angle in the bulk of the triangulation.

The breaking of diffeomorphism symmetry  has considerable repercussions for discrete quantum gravity approaches such as Regge calculus and spin foams \cite{BahrDittrich1,Ditt12Review}. 

In a canonical quantization, diffeomorphism symmetry leads to constraints. A long-standing problem has been to provide an anomaly free representation of this constraint algebra in the quantum theory. The breaking of diffeomorphism symmetry by  discretization, which is often used as a regulator, leads to inconsistent constraints. An alternative formulation, that of `consistent discretizations' \cite{GambiniPullin}, is what we used here.  In this framework broken gauge symmetries lead to pseudo-constraints, which are equations of motion that weakly couple the canonical data of neighboring time slices. (Constraints are equations of motion that involve the data of only one time.) The replacement of the constraints by pseudo-constraints  means that one has more propagating degrees of freedom than in the continuum. Degrees of freedom that are gauge in the continuum are now physical. For example, in  Regge calculus the position of the vertices in the embedding spacetime become physical if vertex translation symmetry is broken. 

In the covariant formalism, breaking diffeomorphism symmetry leads to an unwanted dependence on the choice of triangulation.  In fact, in  \cite{Improved,BahrDittrichSteinhaus} it was conjectured that diffeomorphism symmetry and triangulation invariance are equivalent and shown that diffeomorphism symmetry implies triangulation invariance for one-dimensional systems. Restoring diffeomorphism symmetry is crucial in order to remove discretization or regulator dependence \cite{Ditt12Review}.

To regain diffeomorphism symmetry the work \cite{Improved} suggested the construction of an improved dynamics by coarse graining.  The key point here is that, given fixed boundary data,\footnote{Refinement and coarse graining of the boundary also plays an important role in the coarse graining process \cite{Ditt12}.} refinement of the triangulation leads to smaller deficit angles as the fixed total curvature is distributed over more simplices. Diffeomorphism symmetry is then violated to a lesser extent, and potentially restored, for an infinitely fine triangulation. One constructs an effective action for a coarser triangulation by taking into account the dynamics of the finer one. As the coarse lattice now reflects the dynamics of the refined one, it also exhibits the same amount of diffeomorphism symmetry. Thus one can hope to restore diffeomorphism symmetry for the effective action in the infinite refinement limit.

    This was illustrated successfully in \cite{Improved} using the example of three-dimensional Regge calculus with a cosmological constant. Starting from a discretization with flat simplices, in which diffeomorphism symmetry is broken, the coarse graining procedure yielded as its fixed point an action describing simplices with homogeneous curvature. This action features diffeomorphism symmetry, is triangulation invariant, and leads to an anomaly free constraint algebra \cite{Improved,BonzomDittrich}. 

This has triggered the development of a program for coarse graining spin foam models \cite{Ditt12Review,CoarseRenorm1,CoarseRenorm2,CoarseRenorm3,CoarseRenorm4,CoarseRenorm5,CoarseRenorm6,CoarseRenorm7,HyperCubeCoarse}.  Here the hope is to construct amplitudes for which the regulator dependence is removed and that explicitly display diffeomorphism symmetry. As discussed above, one dynamical quantity that leads to a breaking of diffeomorphism symmetry is curvature, \cite{BahrDittrich1}. Spin foam models so far seem to display non-metricity (Section \ref{introduction}). The question we answer here is whether non-metricity can also lead to breaking of diffeomorphism symmetry.  That it does could be expected due to the connection between triangulation (non-)invariance and (breaking of) diffeomorphism symmetry, and the fact that we found that  the area Regge action is not invariant under two of the Pachner moves. Below we show explicitly that diffeomorphism symmetry is broken for non-metric solutions.  This is important for spin foams as it necessitates understanding the dynamics of the  non-metric degrees of freedom. It also explains the breaking of diffeomorphism symmetry and the triangulation dependence of the spin foam models conjectured to admit no curvature degrees of freedom. We explore the implications of these findings in more detail in the discussion, Section \ref{discussion}.

\subsection{Constructing a non-metric solution}\label{NGsolution}

To analyze diffeomorphism symmetry for a non-metric solution we have to construct these solutions explicitly.  We then evaluate the Hessian on such a background and check whether there are any vanishing eigenvalues.  We could, for instance, consider two consecutive 4-valent tent moves such that we have a bulk vertex at the intermediate time step.  But, there is a short cut we can exploit: in Section \ref{4--valent move}, where we analyzed the 4-valent tent moves, we showed that null vectors for the bulk Hessian lead to null vectors for the non-diagonal-in-time block of the Hessian for the piece of  triangulation that is glued onto the initial triangulation. This piece of triangulation coincides with the initial configuration of the 4--2 Pachner move. It is therefore sufficient to consider this initial Pachner configuration and to evaluate a certain part of the Hessian of the associated action to see if any of its eigenvalues vanish.

To construct a solution with a non-metric area configuration we must first construct non-metric boundary data for the initial 4--2 Pachner move configuration.
To this end we consider the matrix of derivatives $\Gamma^t{}_e = \partial A_t/\partial L_e$. Recall that this matrix identifies the vector space of non-metric area directions. Restricting to the set of boundary areas and boundary edge lengths of the triangulation, the left null vectors $n^I_{\rm bdry}$ of the corresponding matrix $(\Gamma^t{}_e)_{\rm bdry}$ describe the boundary area non-metric directions. Here $I$ labels which null vector is being considered. 

Starting from a metric set of boundary areas $A_t$, we generate a set of non-metric boundary areas by adding multiples of these null vectors  
\be\label{NGArea}
A_{\rm bdry} \rightarrow  A_{\rm bdry}^\kappa = A_{\rm bdry} + \kappa_I \cdot n^I_{\rm bdry},
\ee
with $\kappa_I$  arbitrary, but small parameters. 
The set of areas $A_{\rm bdry}^\kappa$, for non-zero $\kappa_I$, are non-metric in the sense that the corresponding edge lengths are not well defined (i.e.,  the length of a single edge will have different values depending on which simplex it is computed from).

Having fixed the non-metric areas  \eqref{NGArea}, we use them to construct a solution to the equations of motion. 
Here we use the equations of motion \eqref{EOM2} derived from the first order action \eqref{action2}.  These are solved for the dihedral angles $(\theta^\sigma_t)^\kappa$, a Lagrange multiplier $\Lambda_\sigma^\kappa$ for each four--simplex $\sigma$, and the bulk area variables $A_{\rm bulk}^\kappa$. By construction, the set of dihedral angles computed in this way will automatically be compatible with the areas and will satisfy the closure condition for each simplex $\sigma$. The equations of motion impose flatness for each of the bulk triangles. 

\begin{figure}[ht!]
    \centering
    \begin{subfigure}[b]{0.475\textwidth}
        \includegraphics[width=\textwidth]{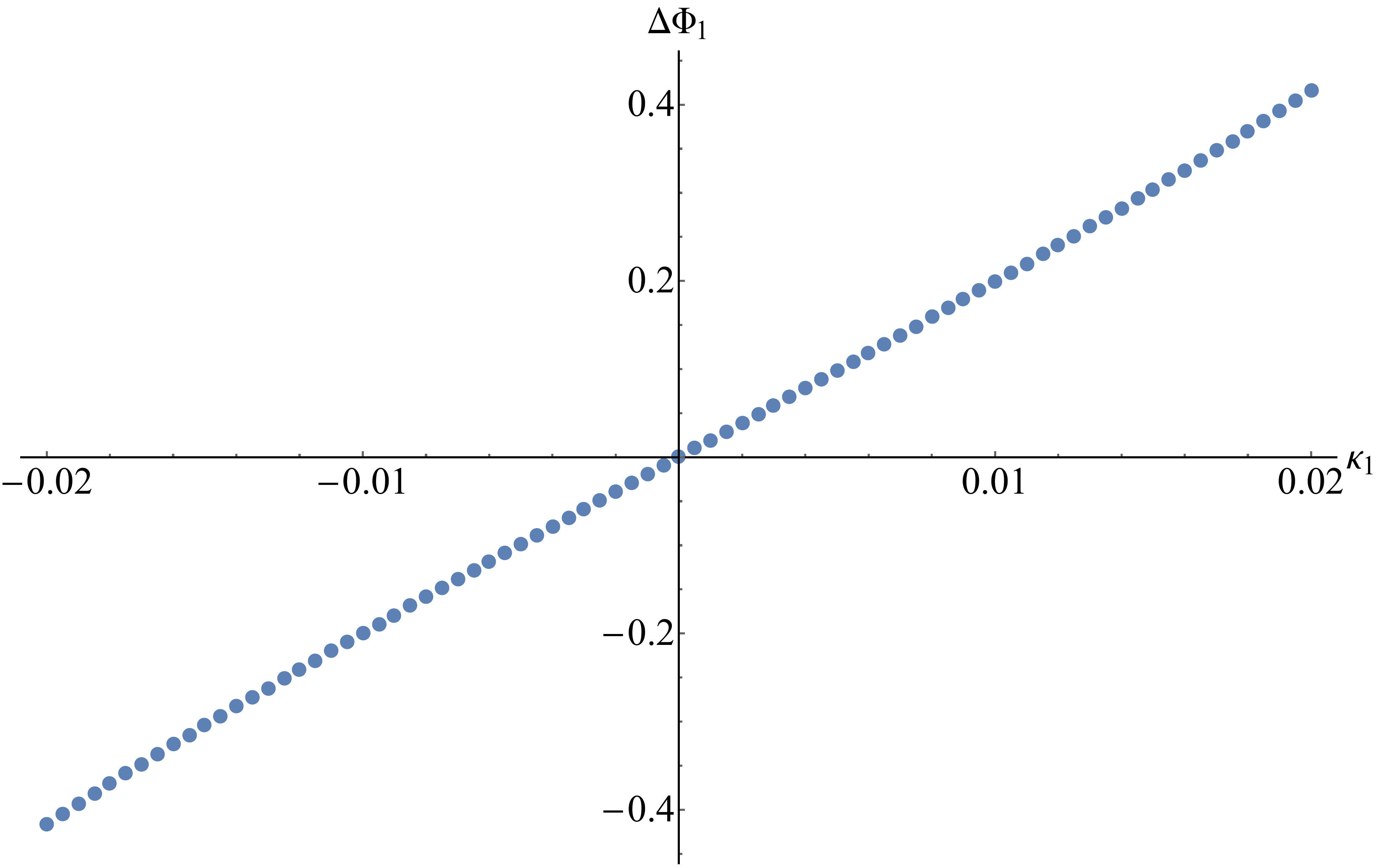}
        \caption{$\Delta\Phi^\kappa_1$ vs. $\kappa_1$ at fixed $\kappa_2=0$.}
        \label{Phi1k1}
    \end{subfigure}
    \qquad 
    \begin{subfigure}[b]{0.475\textwidth}
        \includegraphics[width=\textwidth]{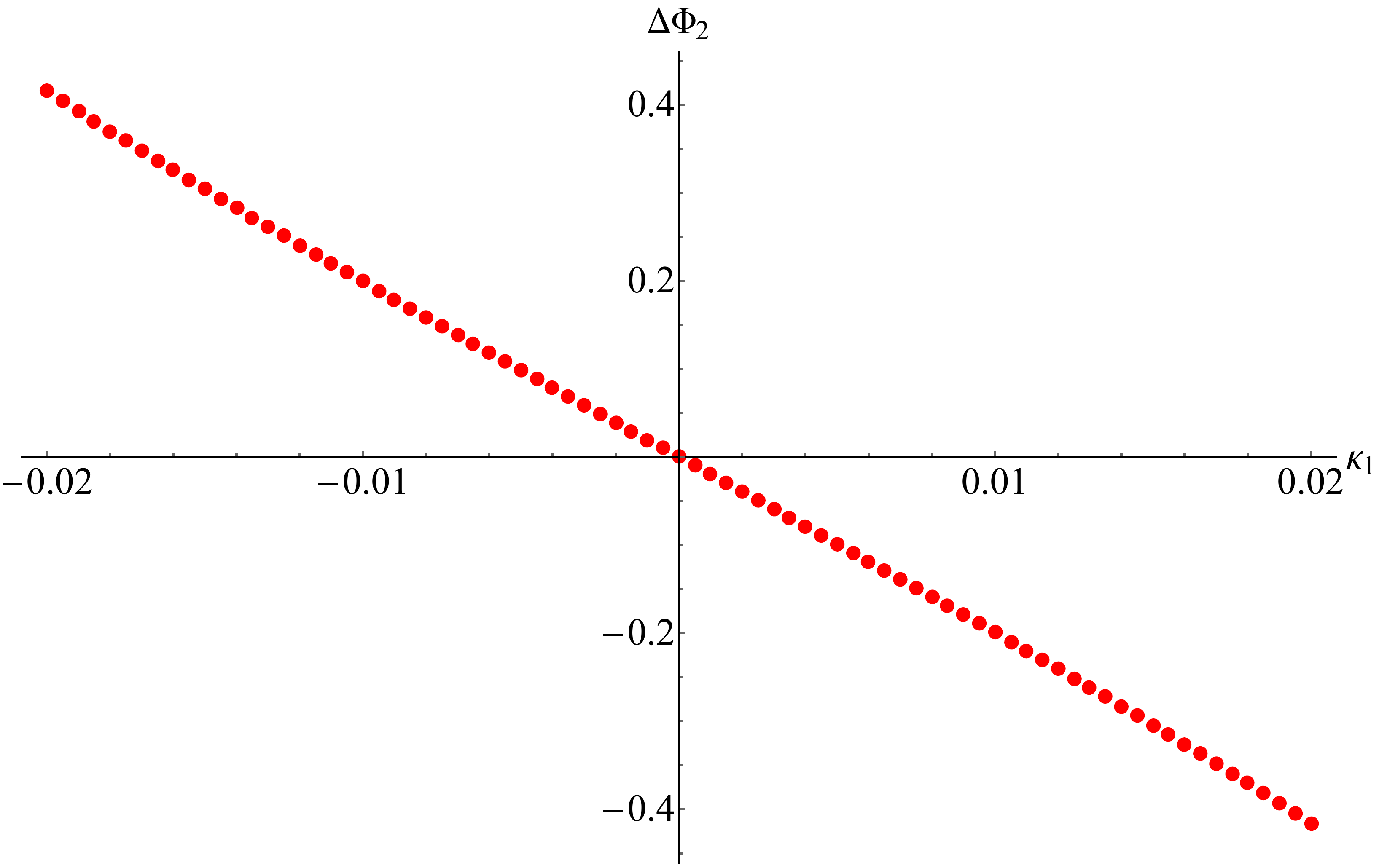}
        \caption{$\Delta\Phi^\kappa_2$ vs. $\kappa_1$ at fixed $\kappa_2=0$.}
        \label{Phi2k1}
    \end{subfigure}
    ~ 
    \begin{subfigure}[b]{0.475\textwidth}
        \includegraphics[width=\textwidth]{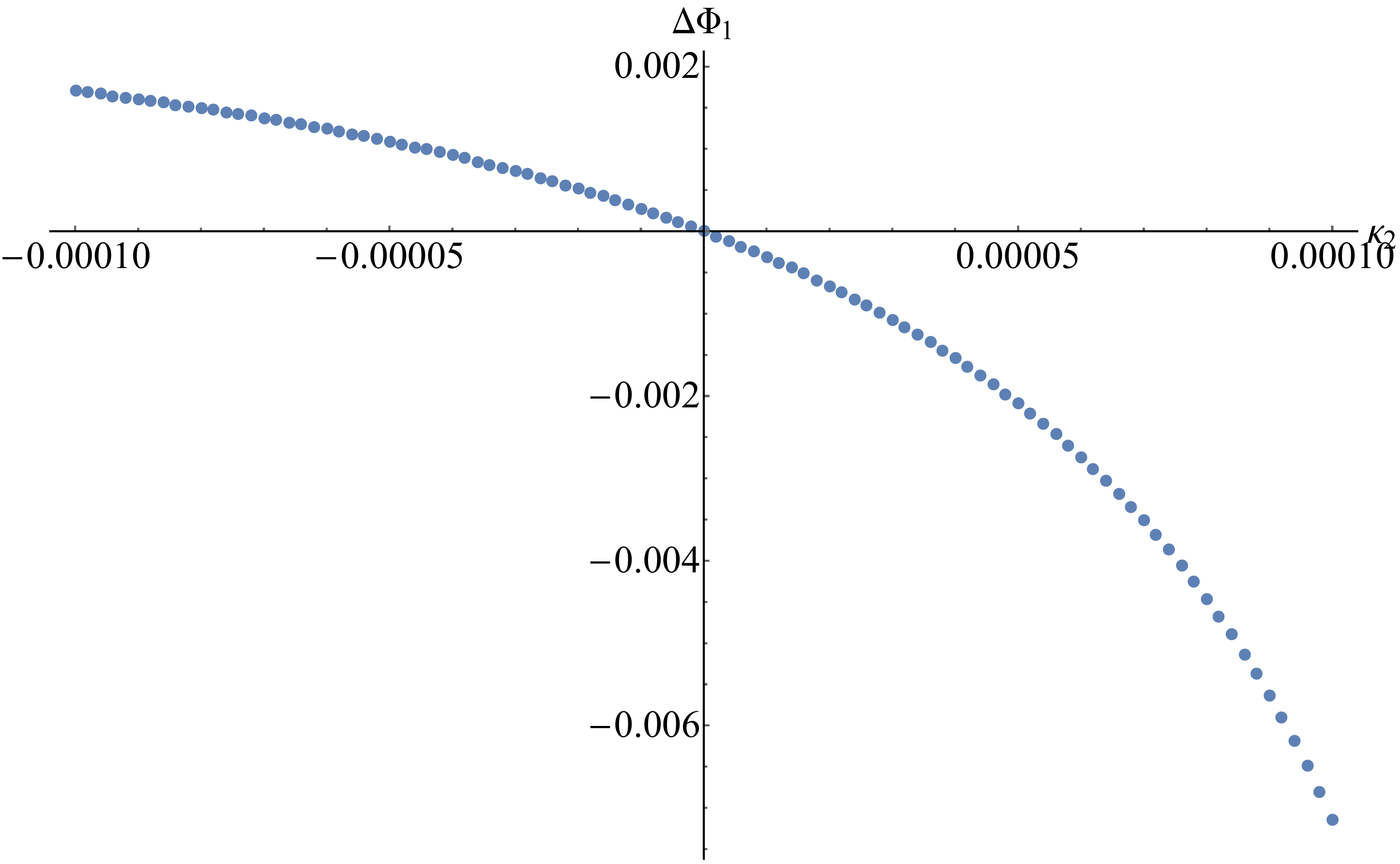}
        \caption{$\Delta\Phi^\kappa_1$ vs. $\kappa_2$ at fixed $\kappa_1=0$.}
        \label{Phi1k2}
    \end{subfigure}
    \qquad
    \begin{subfigure}[b]{0.475\textwidth}
        \includegraphics[width=\textwidth]{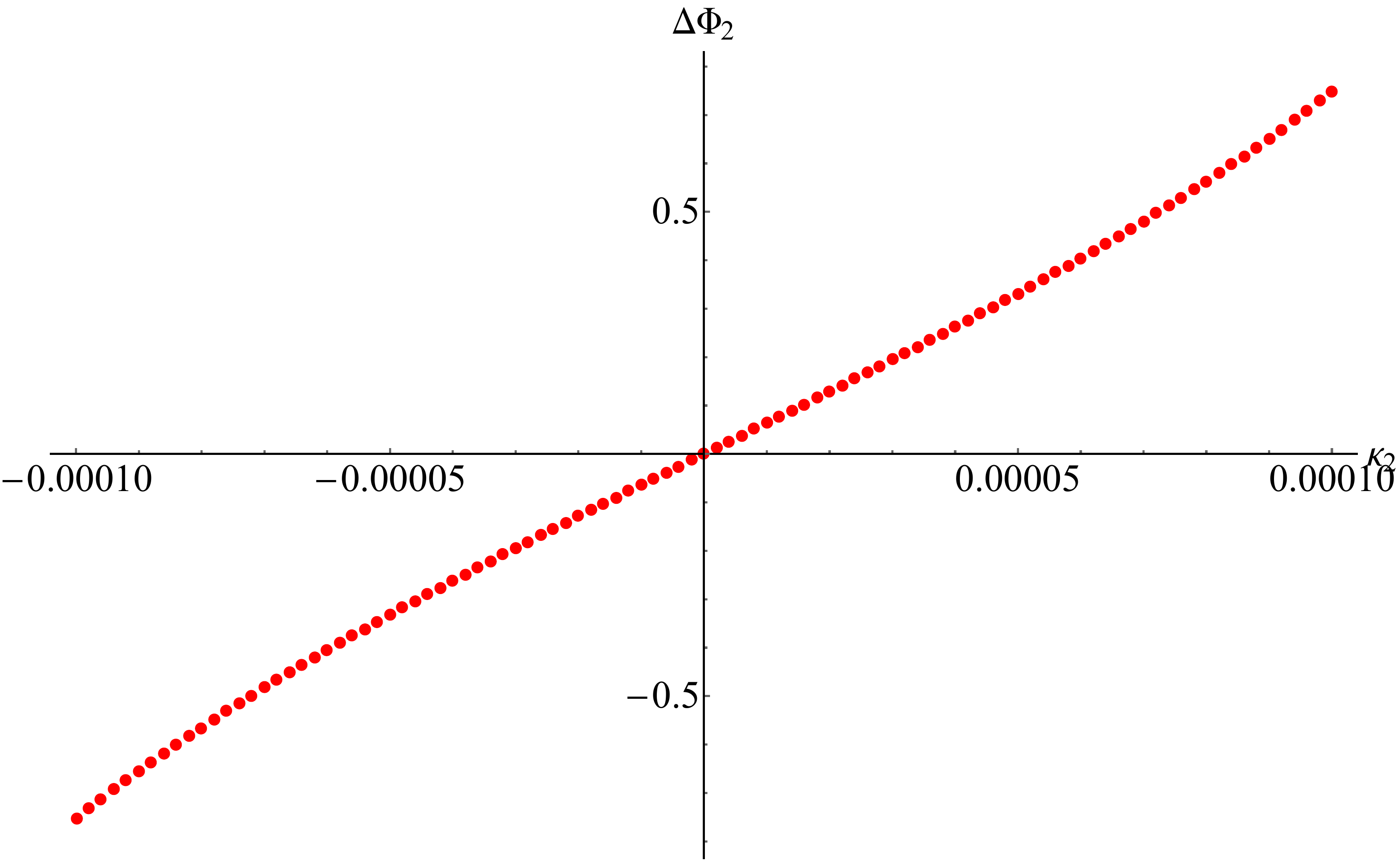}
        \caption{$\Delta\Phi^\kappa_2$ vs. $\kappa_2$ at fixed $\kappa_1=0$.}
        \label{Phi2k2}
    \end{subfigure}    
    \caption{Differences of the 3D dihedral angles $\Delta \Phi_\alpha^\kappa=\Phi_\alpha^\kappa(\sigma^0)-\Phi_\alpha^\kappa(\sigma^1)$  of  tetrahedron $\tau (2,3,4,5)$ computed from the simplices $\sigma^0$ and $\sigma^1$ as functions of the non-metricity  parameters $\kappa_I$. }\label{Phikappa}
\end{figure}

We consider the initial configuration of the  4--2 Pachner move (see Fig. \ref{4-Valent move}). The boundary of this configuration is made up of the two simplices $\sigma^0 = (1,2,3,4,5)$ and $\sigma^1 = (0,2,3,4,5)$, which share the tetrahedron $\tau = (2,3,4,5)$. There are 16  triangles and 14  edge lengths contained in this boundary. As in section \ref{42move}, we consider the variable transformation (with $i,j,k \in \{2,3,4,5\}$ and $\alpha \in \{1,2\}$) 
\be\label{NGTran}
\left( \{ A_{0ij}^\kappa \}, \{ A_{1ij}^\kappa \}, \{ A_{kij}^\kappa \} \right) \rightarrow \left( \{ L_{0i}^\kappa \}, \{ (\Phi^0_{\alpha})^\kappa \}, \{ L_{1i}^\kappa \}, \{ (\Phi^1_{\alpha})^\kappa \}, \{ A_{kij}^\kappa \} \right)\q ,
\ee
 but now for the fully non-perturbative variables (see Appendix \ref{lengthTOareasangles}). Here $ (\Phi^{0}_{\alpha})^\kappa$  and $ (\Phi^{1}_{\alpha})^\kappa$  are the 3D dihedral angles for any choice of two adjacent edges in the tetrahedron shared by the two simplices in the final configuration of the 4--2 Pachner move, the first viewed from $\sigma^1$ and the second from $\sigma^2$.   The difference between these dihedral angles, $\Delta \Phi_\alpha^\kappa := \left( \Phi^{1}_\alpha-\Phi^{0}_\alpha \right)^\kappa$, will make non-metricity transparent (c.f. Fig. \ref{Phikappa}). 

We have chosen an asymmetric, metric background configuration with boundary edge lengths $L_{1i} = L_{04} = L_{05} = 1$, $L_{02} = \sqrt{\frac{8}{9}}$, and   $L_{03} = \sqrt{\frac{9}{8}}$, with $i \in \{2,3,4,5\}$. The matrix $(\Gamma^t{}_e)_{\rm bdry}$ for this configuration has two left null vectors $n_{\text{bdry}}^I, (I  \in \{1,2\})$, the non-metric directions for these boundary areas. Using these null vectors, we construct the non-metric areas $A_t^\kappa$ using \eqref{NGArea} and then transform them to length and 3D dihedral angles, as in Eq. \eqref{NGTran}. 

In Figure \ref{Phikappa}, we have plotted the two parameters $\Delta \Phi_\alpha^\kappa$ against $\kappa_I$ for the edges $(24)$ and $(25)$. Fixing $\kappa_2 =0$, $\Delta \Phi_1^\kappa$ increases linearly with $\kappa_1$ while  $\Delta \Phi_2^\kappa$ decreases, see panels  (a) and (b). On the other hand, fixing $\kappa_1 = 0$,  $\Delta \Phi_2^\kappa$ grows monotonically with $\kappa_2$ and $\Delta \Phi_1^\kappa$ decreases monotonically, panels (c) and (d). 

\subsection{Breaking of diffeomorphism symmetry}

Having produced a non-metric boundary configuration,  we numerically  solve for the bulk variables ($A_{\rm bulk}^\kappa, (\theta^\sigma_t)^\kappa,\Lambda_\sigma^\kappa$). These bulk variables belong to the configuration consisting of the four simplices $\sigma^2,\sigma^3,\sigma^4$, and $\sigma^5$ that share the bulk edge (01). We can now evaluate the Hessian of the area Regge action on these solutions, and, as in Section \ref{4--valent move}, compute an effective, linearized action for the 4-valent tent move, but now on a  non-metric background.  We find that the mixed time block of the effective Hessian  is not singular for non-vanishing non-metricity parameters. In particular, all eigenvalues of the mixed time block of the Hessian are non-vanishing.\footnote{Using the highly symmetric background  from Section \ref{42move} as a starting point, one finds that 1 of the 4 eigenvalues, which vanished on a metric background, remains zero under a deformation to a non-metric boundary. Here we have a less symmetric configuration but, as it is close to the very symmetric background, one of the eigenvalues is growing slowly compared to the others.} The gauge symmetries (vertex translation) are therefore broken by the non-metric boundary data.

\begin{figure}[!h]
   \centering
    \begin{subfigure}[b]{0.485\textwidth}
        \includegraphics[width=\textwidth]{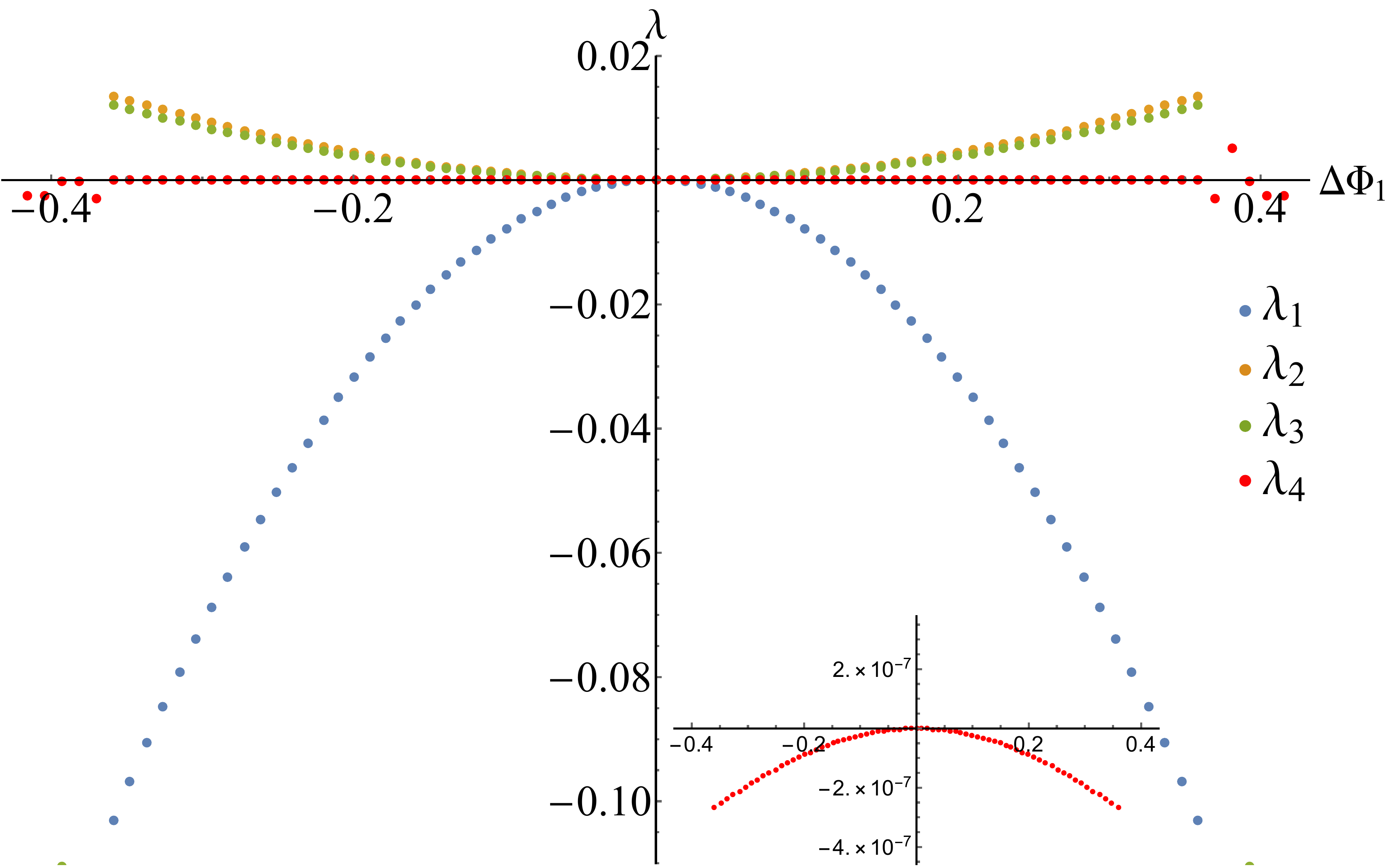}
        \caption{The smallest eigenvalues as a function of  $\Delta\Phi^\kappa_1$ with $\kappa_1=0.$}
        \label{LaPhi1k1}
    \end{subfigure}
    \quad 
    \begin{subfigure}[b]{0.485\textwidth}
        \includegraphics[width=\textwidth]{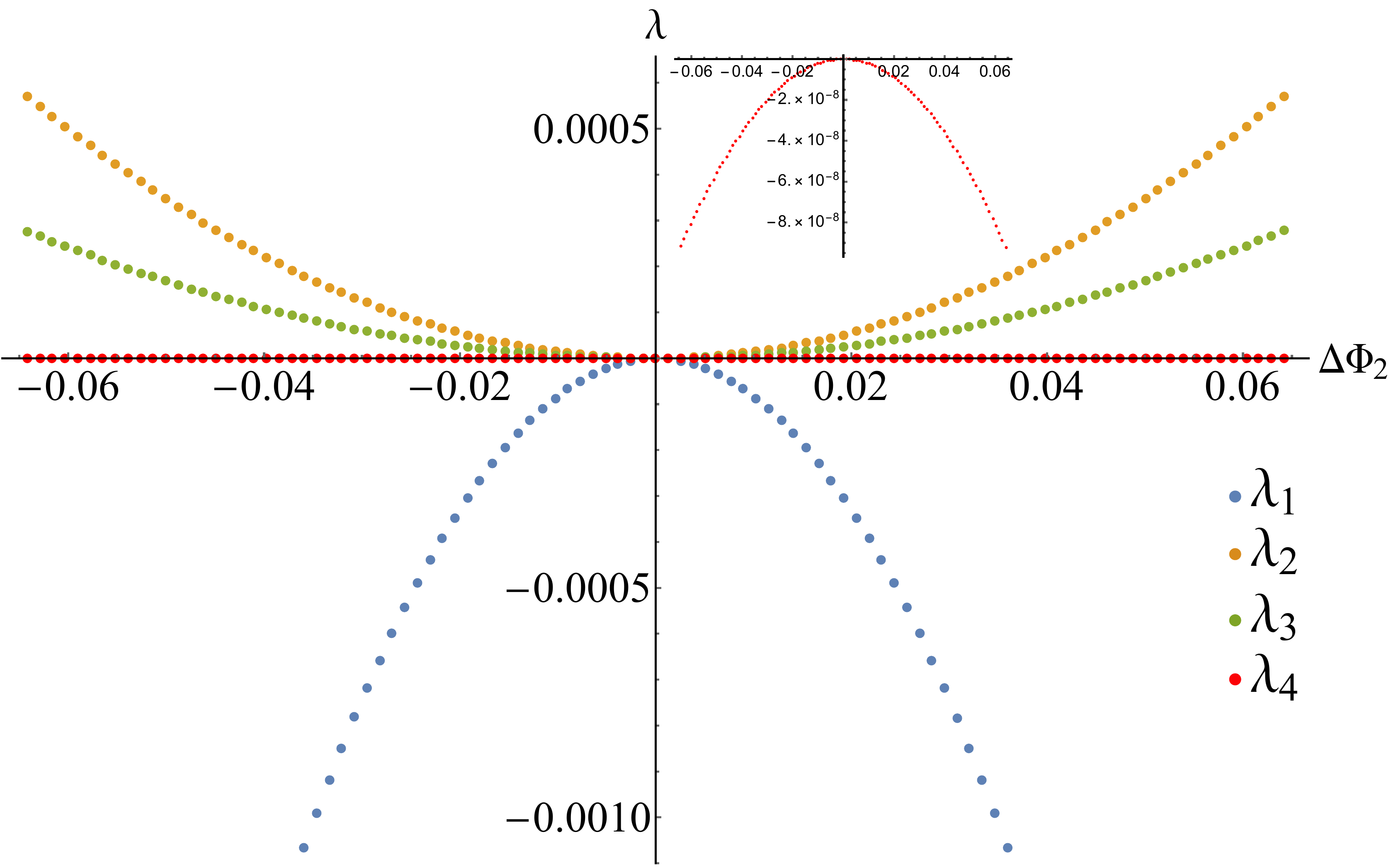}
        \caption{The smallest eigenvalues as a function of $\Delta\Phi^\kappa_2$ with $\kappa_2=0.$}
        \label{LaPhi2k2}
    \end{subfigure}    
    \caption{Eigenvalues of the mixed time block of Hessian as a function of  $\Delta\Phi_\alpha^\kappa$. The lowest eigenvalues are zoomed in at the bottom right corner of (a)  and the top right corner of (b) . }\label{LambdaPhi}
\end{figure}

The panels of Figure \ref{LambdaPhi} show the four smallest eigenvalues 
as a function of the non-metricity parameters $\Delta\Phi_{\alpha}^\kappa$.
We observe that all these eigenvalues, including the lowest eigenvalue $\lambda_4$ (shown with a separate scale), 
grow quadratically with the non-metricity parameters. 
Panel (a) is plotted at fixed $\kappa_1=0$, while panel (b) is for $\kappa_2=0$. Similar behaviors appear for the other two combinations of non-metricity parameter and $\kappa_I$. All the eigenvalues vanish identically only in the metric case $\Delta\Phi_{1}^\kappa = \Delta\Phi_{2}^\kappa = 0$ (or $\kappa_I = 0$). In practice the eigenvalues are computed numerically and never exactly vanish; at $\kappa_{I}=0$ the two non-vanishing eigenvalues are seven orders of magnitude larger than the largest `vanishing' eigenvalue. 

We thus can conclude that the vertex translation symmetry, which is present on metric backgrounds, is broken for non-metric backgrounds. In the examples we considered here, the  relevant Hessian eigenvalues grow quadratically with our non-metricity parameter, the difference of 3D dihedral angles seen from two different 4-simplices. This is similar to the findings on diffeomorphism breaking in length Regge calculus \cite{BahrDittrich1}, where the eigenvalues also grew quadratically with one of the deficit angles in the bulk triangulation.

\section{Discussion}\label{discussion}

With the aim of reaching a better understanding of spin foam dynamics, we have revisited area Regge calculus. We have provided a well-defined action principle for flat, as well as homogeneously curved, simplices and analyzed certain aspects of the covariant dynamics, in particular, the behavior of area Regge calculus under Pachner moves. The invariance properties of area Regge calculus under these moves differ from those of length Regge calculus. Obtaining a semiclassical limit of Pachner moves seems feasible for spin foams \cite{BanburskiEtAlPachner,RielloPachner1,RielloPachner2,ChenPachner}, and so our results can be used as a test to differentiate between the different types of dynamics in spin foam models. 

Interestingly, the equations of motion can impose flatness in the 3--3 move even when boundary data would seem to induce curvature. This is due to the surprising fact that, although there are more area than length variables in the boundary, the boundary areas do not always uniquely determine the boundary lengths. 

We have also performed a canonical analysis of area Regge calculus using tent moves. For the linearized dynamics over a metric background we find the constraints resulting from the diffeomorphism symmetry of the (linearized) action. The same constraints arise for (linearized) length Regge calculus on a flat background. We have not found additional constraints. As there are generically far more areas than lengths, area Regge calculus has far more physical degrees of freedom then length Regge calculus. In particular, for an $n$--valent tent move we expect $3n-10$ physical degrees of freedom in area Regge calculus and $n-4$ physical degrees of freedom in length Regge calculus. We have provided an in-depth analysis of how the non-metric degrees of freedom appear and discussed how they can be parametrized in the 4-valent tent move. Our results suggest that the differences of 3D dihedral angles as determined from different four-simplices is a good measure for the non-metricity in general.

We analyzed the gauge symmetry content of area Regge calculus and found that on metric backgrounds area Regge calculus features (discrete remnants of) diffeomorphism symmetry. These symmetries are broken if one considers non-metric backgrounds. The breaking can be quantified via the size of the eigenvalues of the Hessian evaluated on these backgrounds. There is a  quadratic dependence on our non-metricity parameter, the difference of certain 3D dihedral angles.

This makes area Regge calculus an interesting model for testing how to regain diffeomorphism symmetry via coarse graining. Area Regge calculus is also a credible candidate for describing the semiclassical limit of the Barrett-Crane spin foam model \cite{BarrettCrane}---such a test can thus be extended to a non-perturbative quantum theory.  Recovered diffeomorphism symmetry is conjectured to lead to triangulation invariant models. Thus, if a meaningful  continuum limit can be reached, one might uncover interesting topological invariants for four-dimensional manifolds based on generalized geometries defined by an area measure. 

This opens up the question of what kind of continuum limit area Regge calculus might have. There are various proposals for continuum (quantum gravity) theories in which the areas are the fundamental variables \cite{Schuller:2005yt, Raamsdonk,Raamsdonk2, RyuTak}, and the question is whether these can be connected to area Regge calculus. 

Let us now discuss the implications of our findings for spin foams.  We have investigated area Regge calculus because it is arguably the simplest theory that features non-metric degrees of freedom in the form of non-shape matching of triangles that are glued to one another. Similar non-metricities are also suspected to appear in spin foams, but their role in the dynamics of the models is an open question.  Here we showed that such non-metricities can lead to a different behavior under Pachner moves. A semiclassical analysis of such moves can therefore reveal which kind of dynamics---that of length Regge calculus or that of area Regge calculus---is implemented in the various spin foam models. Note that here we analyzed only the classical action. Although the classical action of length Regge calculus is invariant under 5--1 and 4--2 moves one can prove that even in the length calculus there is no local measure for a state sum that is invariant under these moves \cite{DittKaminStein}. One might therefore need a semiclassical analysis. 

The findings here also provide several cautionary notes for the coarse graining program for spin foams \cite{Ditt12Review,CoarseRenorm1,CoarseRenorm2,CoarseRenorm3,CoarseRenorm4,CoarseRenorm5,CoarseRenorm6,CoarseRenorm7,HyperCubeCoarse}. Firstly, finding a restoration of diffeomorphism symmetry and propagating degrees of freedom in the continuum limit will not be sufficient to prove that spin foams have general relativity as continuum limit. We will also need to understand whether these propagating degrees of freedom are metrical or instead describe non-metric degrees of freedom. Unfortunately, there seem to be much more of these non-metric degrees of freedom than metrical ones, but the detailed counting will depend on the specific model. 

Secondly, we have found that not only curvature, but also non-metricity leads to a breaking of diffeomorphism symmetry. Whereas there are arguments that under refinement the curvature per simplex gets smaller, and thus diffeomorphism symmetry may be restored, we do not know of such arguments for the non-metric degrees of freedom. The works \cite{HyperCubeCoarse} consider a setup where the degrees of freedom for the spin foams are drastically reduced so that {\it only} non-metric degrees of freedom remain.  They show that a certain global remnant of the vertex translation symmetry can be restored under coarse graining.\footnote{These symmetries also affect the boundary and are thus technically not gauge symmetries.} This is already a good sign. But, the reduction of degrees of freedom  in \cite{HyperCubeCoarse}  also drastically affects  the number of non-metric degrees of freedom. In these works they scale with the linear size of the lattice instead of with its volume. Finally, for these studies the action vanishes and only one measure parameter is changed under coarse graining, whereas our discussions are focussed on the action. 
Therefore a crucial task will  be to better understand the non-metric degrees of freedom in spin foams and, in particular, whether their number dominates over the metric degrees of freedom, as this could impede restoration of diffeomorphism symmetry. 

This motivates a number of future research directions. Firstly, it would be very helpful to understand the dynamics of area Regge calculus on (regular) lattices involving a large number of simplices. In particular, the question is whether the non-metric degrees of freedom  show wave-like propagation, perhaps along lines similar to \cite{BarrettRefWaves}. Also it would help to establish whether the non-metricity at each simplex, e.g. the difference of 3D dihedral angles used in this work, decreases under refinement. This would make the scenario where diffeomorphism symmetry is restored in theories with non-metricity in the continuum limit much more viable. 

 The dynamical evolution of non-metric degrees of freedom over more time steps could be studied in a simple setup using `cylinder moves'. The idea we have in mind for a cylinder moves is to focus on the three-dimensional star of a vertex $v$. Starting with a tent move at the vertex $v$ one then glues more simplices on top of the tent so that one obtains a cylinder with a finite time elapsed also on the boundary.  Such moves could be iterated and thus one could study the propagation of bulk degrees of freedom in a setup where a finite time passes both in the bulk and at the boundary of the cylinder.

Secondly, area Regge calculus is only one model showing non-metric degrees of freedom. It is a credible candidate for describing the semiclassical limit of the Barrett--Crane model but maybe less so for the newer spin foam models \cite{NewSFM1,NewSFM2,NewSFM3,NewSFM4,NewSFM5}. Here the action proposed in \cite{AreaAngle} seems to be a better candidate. It is based on area variables and 3D dihedral angles. The action in \cite{AreaAngle} adds two types of constraints to this action so that the theory becomes equivalent to length Regge calculus. The first type are the so--called closure constraints and the second are gluing or (triangle) shape matching constraints. The second type of constraints is not implemented in loop quantum gravity \cite{DittRyan1,DittRyan2} and its status in spin foam models is unclear.   A similar action has also been identified in \cite{Hnybida} (without the shape matching constraints) as the semiclassical limit of  a simplex amplitude built from coherent intertwiner states.

It would be helpful to analyze the area-angle Regge action \cite{AreaAngle}, but with the shape matching constraints removed or weakened. The first question would be to construct a well-defined action, the second to determine the dynamics and the types of physical degrees of freedom. The main questions are whether this action allows for a dynamics with curvature and how the non-metric degrees of freedom propagate.

The present work shows that there might be a variety of models based on various geometric variables that are related to Regge calculus and candidates for spin foam models. We have suggested some ways in which to use these models to better understand  the dynamics of spin foams, e.g. by comparing the behavior of these models under Pachner moves. A systematic understanding of the dynamics and kinematics of generalized geometries will not only help to understand the dynamics of spin foam models, but also, if necessary, to improve these models.

\section*{Acknowledgements}

HMH is grateful to Simone Speziale for discussion. HMH thanks the IGC at Pennsylvania State University for warm hospitality while completing this work and the Perimeter Institute for Theoretical Physics for generous sabbatical support. SKA is supported by an NSERC grant awarded to BD. This work is  supported  by  Perimeter  Institute  for  Theoretical  Physics.   Research  at  Perimeter  Institute is supported by the Government of Canada through Industry Canada and by the Province of Ontario through the Ministry of Research and Innovation.

\appendix

\section{Curved simplex areas as functions of the dihedral angles}
\label{curvedanalytics}

We use the drop vertex notation, where $\sigma(k)$ is the tetrahedron obtained from the four-simplex $(ijklm)$ by dropping the vertex $k$ and $\sigma(kl)$ is the triangle obtained by dropping vertices $k$ and $l$. This triangle contains edges $\sigma(ikl)$ and $\sigma(jkl)$. Denote the 2D face angle between these two edges by $\alpha_{ij,kl}$, where the second index pair indicates the triangle $\sigma(kl)$ and the first pair indicates the vertices dropped to obtain each of the edges. Similarly, the 3D dihedral angle in tetrahedron $\sigma(k)$ between triangles $\sigma(ik)$ and $\sigma(jk)$ and hinged at the edge $\sigma(ijk)$ is denoted $\phi_{ij,k}$. Finally, the 4D dihedral angle between the tetrahedra $\sigma(i)$ and $\sigma(j)$ and hinged at $\sigma(ij)$ is $\theta_{ij}$. 

Because all of these angles are defined in the appropriate tangent space many of the properties of flat simplices can be carried over to the spherical case. For example, by intersecting a small enough neighborhood of the vertex $m$ (contained in simplex $(ijklm)$) by a sphere, the spherical cosine law yields
\be
\cos \alpha_{ij, kl} = \frac{\cos \phi_{ij,k} + \cos\phi_{il,k} \cos \phi_{jl,k}}{\sin \phi_{il,k} \sin \phi_{jl,k}}.
\ee
Carrying out the analogous argument in one higher dimension also yields
\be
\cos \phi_{ij,k} = \frac{\cos \theta_{ij} + \cos \theta_{ik} \cos \theta_{jk}}{ \sin \theta_{ik} \sin \theta_{jk}}. 
\ee
Using these two results we can relate the 2D face angles and the 4D dihedral angles. First note that all of these dihedral angles take values in the range $\theta \in [0,\pi]$ and we can safely exchange $\sin \theta = \sqrt{1-\cos^2 \theta}$. Then after a little algebra, and briefly adopting the shorthand $ \cos \theta \equiv \c \theta$, we have
\be
\c \alpha_{ij, kl} = \frac{\c \theta_{ij} +\c \theta_{ik} \c \theta_{jk}+\c \theta_{il} \c \theta_{jl} -\c \theta_{ij} \cs \theta_{kl}+\c \theta_{il} \c \theta_{jk}\c \theta_{kl} +\c \theta_{jl} \c \theta_{ik}\c \theta_{kl}}{\sqrt{(1- \cs \theta_{il}- \cs \theta_{ik}- \cs \theta_{kl}-2\c \theta_{il} \c \theta_{ik}\c \theta_{kl})(1- \cs \theta_{il}- \cs \theta_{ik}- \cs \theta_{kl}-2\c \theta_{il} \c \theta_{ik}\c \theta_{kl})}}.
\ee
Combining this result with the spherical excess formula for the area of a triangle, $A_t = \alpha+\beta+\gamma - \pi,$ gives a general expression for $A_t(\theta^{\sigma})$. With the appropriate change to the formula for the area, the same results apply to a finite simplex in the hyperbolic case. 


\section{The Gram matrix and the derivatives of its determinant}\label{SimplexK}

Consider an  $n$-simplex $\sigma$. It has $(n+1)$ vertices $v_i $, $i \in \{ 1\cdots n+1\}$ and $(n+1)$ faces $f_i$ (defined as the $(n-1)$-simplex obtained by removing the vertex $v_i$ from the simplex $\sigma$), each face with a corresponding volume $V_i$. Let $\hat{n}_i$ be the outward unit normal to the face $f_i$. Each pair of faces $f_i$ and $f_j$ share a common hinge $h_{ij}$ (the $(n-2)$-simplex obtained by removing the vertices $v_i$ and $v_j$) and the angle between the unit normals $\hat{n}_i$ and $\hat{n}_j$ defines the dihedral angle at the hinge $h_{ij}$ 
\ba
\cos\theta_{ij}\,=\, -\langle \hat{n}_i , \hat{n}_j \rangle \q,
\ea
where $\theta_{ij}$ is the dihedral angle between the faces $f_i$ and $f_j$. The {\it Gram matrix} is defined as the symmetric matrix $G^\sigma$ whose elements are given by 
\[  
G^\sigma_{ij} =  -\cos(\theta^\sigma_{ij}) \q ,\qquad G^\sigma_{ii} = 1.  
\]
Every closed flat simplex $\sigma$ satisfies the closure constraint 
\ba\label{closure}
\sum_i V_i \,\hat{n}_i = 0 \q,
\ea
where $V_i$ is the volume of the face $f_i$. Using \eqref{closure}, the Gram matrix has a null space given by the vector with entries $\{V_i\}$, i.e.
\ba \label{cclos}
\sum_{j} G^\sigma_{ij}V_j\,  = \, \sum_{j} \langle \hat{n}_i , \hat{n}_j \rangle V_j \,=\,  \sum_{j} \langle \hat{n}_i , V_j\,\hat{n}_j \rangle  \, =\, 0  .
\ea
It follows that the Gram matrix is singular and  $\det (G^\sigma) = 0$. Any other null vector of $G^\sigma$ is proportional to $V_i$, therefore $V_i$ is the only null vector for $G^\sigma$.  For a non-degenerate simplex the $(n+1)$  vectors $\hat n_i$ span an $n$-dimensional space, thus there are no further null vectors for the Gram matrix in this case. 

Using Jacobi's formula for matrices, we can express the derivative of the determinant of the Gram matrix in terms of its adjugate matrix  $\text{adj}(G^\sigma)$:
\ba\label{Jacobi}
d \det(G^\sigma) = \text{Tr} \left( \text{adj}(G^\sigma)\, dG^\sigma \right) \q ,
\ea
and the second derivative is given by
\ba
dd \det(G^\sigma) = \text{Tr} \left(d\text{adj}(G^\sigma)\, dG^\sigma \right) \,+\, \text{Tr} \left( \text{adj}(G^\sigma)\, ddG^\sigma \right)  \q .
\ea
The adjugate matrix of any matrix $A$ is defined as the transpose of the matrix of cofactors of $A$, with the cofactor matrix given by $C_{ij} = (-1)^{i+j}\det A({ij}) $ where $A({ij})$ is the matrix $A$ with the $i^{th}$ row and $j^{th}$ column removed. The adjugate matrix satisfies the relation 
\ba\label{adj}  \text{adj}(A) A = A\, \text{adj}(A) = \det(A)\mathbb{I} \q ,  \ea
where $\mathbb{I}$ is the identity matrix. 

For the second derivative of the Gram matrix we thus need the derivative of the adjugate of the Gram matrix. The usual trick of taking the derivative of equation (\ref{adj}) does not help, as $G^\sigma$ is not invertible. We thus have to use the explicit definition of the adjugate in terms of the determinants of sub-matrices and use again Jacobi's formula
\ba
d(\text{adj}(G^\sigma))_{ij}\,=\, (-1)^{i+j} d \det G^\sigma(ji) \,=\,  (-1)^{i+j} \,\text{Tr} \left( \text{adj}(G^\sigma(ji))\, dG^\sigma(ji) \right) \q .
\ea

Following \cite{RivinMultiSine} we will now determine the structure of the adjugate ot  the Gram matrix. As noted above the Gram matrix for a non--degenerate simplex has exactly one null vector. For a symmetric matrix $A$ which has a unique null vector $N$, the adjugate is given by
\ba 
\text{adj}(A)_{ij} = C \, N_i N_j   \q ,
\ea
with $C$ a constant that we will determine below.

\begin{proof}
Since $A$ has only one null vector and $\det (A) =0$, the relation given in \eqref{adj} implies that the image of $\text{adj}(A)$ is contained in the kernel of $A$, which is given by $N$. Hence $\text{adj}(A)$ has rank 1. As  $A$  and therefore $\text{adj} A$ is symmetric we can conclude that $\text{adj}(A)_{ij} = C \, N_i N_j $.
\end{proof}

By definition, the principal $n$ minors of an $(n+1)\times (n+1)$ matrix $A$ are given by the diagonal elements of $\text{adj}(A)$. If $A$ has only one zero eigenvalue, then from the characteristic polynomial, the product of non-zero eigenvalues equals the sum of the principal $n$ minors of $A$
\ba 
C \sum_i  N_i N_i = \prod_{\lambda_i\neq 0} \lambda_i . 
\ea\\

For the Gram matrix the null vector is given by $N_i=V_i$ and the adjugate is thus $\text{adj}(G^\sigma) = C V_i V_j$. The constant $C$ is given by
\ba
C\,=\, \frac{\prod_{\lambda_i\neq 0} \lambda_i }{\sum_k V_k^2}  \,=\, \frac{ \det( G^\sigma_{ij}+V_iV_j)}{ \left(\sum_k V_k^2\right)^2} \q.
\ea

Therefore for the derivate of the Gram matrix is 
\ba
d \det(G^\sigma) 
\,=\, 2C \,\sum_{i < j} V_i V_j \sin(\theta^\sigma_{ij}) d \theta^\sigma_{ij} \,=\, 2C   \sum_{i < j}   \frac{n}{n-1}  \, V V_{ij}  d \theta^\sigma_{ij}
\ea
where we used the generalized law of sines (see e.g.  \cite{Kokkendorff,DittFreidSpez} for a derivation)
\ba
\sin \theta_{ij}\,=\, \frac{n}{n-1} \frac{V_{ij}V}{V_iV_j} \q, 
\ea
 $V$ is the volume of the simplex $\sigma$, and $V_{ij}$ is the volume of the hinge $h_{ij}$.

We thus have
\ba
\frac{\partial \det(G^\sigma)}{\partial \theta_{ij}}\,=\, c' \, V_{ij}   \; ,   \qquad c'=  2\,\frac{n}{n-1} V \,\frac{ \det( G^\sigma_{ij}+V_iV_j)}{ \left(\sum_k V_k^2\right)^2} \q .
\ea

\section{Edge lengths from areas and dihedral angles of a tetrahedron} \label{lengthTOareasangles}

We derive a formula for the edge lengths of a tetrahedron expressed as a function of its four triangle areas and any two of its dihedral angles along non-opposite edges. As was explained in Appendix \ref{SimplexK},  the Gram matrix has one null vector whose components are given by $V_i$. For a tetrahedron this gives a set of four equations for the 3D dihedral angles $\phi_{ij}$
\ba\label{clos}
\sum_{i\neq j} V_i \cos\phi_{ij}  -V_j \,=\, 0 
\ea
where $V_i$ is the  area of the triangle obtained from $\tau$ by dropping the vertex $i$. 
We can solve for four of the dihedral angles as a function of the four areas and two of the dihedral angles.\footnote{One might wonder why the closure relations (\ref{clos}), which constitute only 3 independent equations for a tetrahedron, allow one to solve for 4 quantities. The reason is that the $\phi_{ij}$ are also not independent; they satisfy $\det G^\tau=0$. 
} The remaining two dihedral angles must be such that they are non-opposite. In general, for an $n$-simplex, $n+1$ dihedral angles can be solved for and $n$ of these must be the dihedral angles of the $n$ hinges of any face $f_i$. As an example, for a tetrahedron $\tau$ with  vertices $(1,2,3,4)$ one can solve for the 3D (interior) dihedral angles $\phi_{ij}$
\ba
\cos \phi_{14} &=& \frac{V_1 - V_2\cos\phi_{12} -V_3\cos\phi_{13} }{V_4}\q, \\
\cos \phi_{23} &=& \frac{V_1^2 +V_2^2+V_3^2 -V_4^2- 2V_1\left( V_2\cos\phi_{12} +V_3\cos\phi_{13} \right) }{2V_2V_3} \q, \\
\cos \phi_{24} &=& \frac{-V_1^2 +V_2^2-V_3^2 +V_4^2 + 2V_1V_3\cos\phi_{13} }{2V_2V_4} \q,\\
\cos \phi_{34} &=& \frac{-V_1^2 -V_2^2+V_3^2 +V_4^2 + 2V_1V_2\cos\phi_{12} }{2V_3V_4} \q .
\ea

Heron's formula gives the area of a triangle as a function of its edge lengths. For the triangle $\tau(i)$, 
\be\label{heron}
V_i = \frac{1}{4}\sqrt{   \sum_{j,k\neq i} V_{ij}^2 V_{ik}^2 - 2 \sum_{j\neq i} V_{ij}^4 } \q, 
\ee
where  $V_{ij}$ is the length of the edge obtained by dropping  the vertices $i$ and $j$ from $\tau$.  We shall also make use of the generalized law of sines in three dimensions 
\be\label{3dsinlaw}
V_{ij} = \frac{2}{3} \frac{V_i V_j \sin\phi_{ij}}{V} \q, 
\ee
with $V$  the volume of the tetrahedron. Using the generalized law of sines and Heron's formula, Eqs. \eqref{heron} and \eqref{3dsinlaw}, one can compute the volume of the tetrahedron as a function of its four areas and all 6 dihedral angles. The result is given by (N.B. $\sin\phi_{ii} =0$)
\be\label{volum}
V = \frac{1}{3} \sqrt[4]{V_i^2 \left( \sum_{j,k} V_{j}^2 V_{k}^2\sin^2\phi_{ij} \sin^2\phi_{ik} - 2 \sum_{j} V_{j}^4 \sin^4\phi_{ij}   \right)} \q .
\ee
Substituting this formula into the generalized law of sines, one obtains  a formula for the edge lengths of a tetrahedron as a function of its four areas and all its dihedral angles. Then, using the four solutions for the dihedral angles from (C2-C5), we get the edge lengths of the tetrahedron as a function of the areas of the face triangles and two non-opposite dihedral angles.

As an example, we compute the edge length $e(34)$ of the tetrahedron $\tau(1,2,3,4)$ as a function of the four areas and the dihedral angles $\phi_{12}$ and $\phi_{13}$. Using the short hand notation $ \sin \phi_{ij} \equiv \s \phi_{ij}$, we have
\be
V_{12} =  \frac{ 2 \sqrt{V_1} V_2 \, \s\phi_{12}}{  \left( 2\left[ V_{2}^2 V_{3}^2\, \s\phi^2_{12} \s\phi^2_{13} + V_{2}^2 V_{4}^2\, \s\phi^2_{12} \s\phi^2_{14} + V_{3}^2 V_{4}^2\, \s\phi^2_{13} \s\phi^2_{14} \right]-  V_{2}^4\, \s\phi^4_{12} -  V_{3}^4\, \s\phi^4_{13} - V_{4}^4 \, \s\phi^4_{14}   \right)^{1/4} },
\ee
where 
\be
\sin^2\phi_{14}  =  1 - \left( \frac{V_1 - V_2\cos\phi_{12} -V_3\cos\phi_{13} }{V_4} \right)^2  \q .
\ee

\end{document}